\documentclass[reprint,amsmath,amssymb,aps,prb,superscriptaddress,longbibliography]{revtex4-2}
\usepackage[T1]{fontenc}
\usepackage{graphicx}
\usepackage{tabularx}
\usepackage{mathrsfs}
\usepackage{bm}
\usepackage{bbm}
\usepackage{hyperref}
\usepackage{dcolumn}
\usepackage{amsmath}
\usepackage{amssymb}
\usepackage{xcolor}
\usepackage{subfig}
\usepackage{caption}
\usepackage{makecell}
\usepackage{siunitx}
\usepackage{ulem}
\usepackage{textcomp}
\usepackage{booktabs}
\usepackage[version=4]{mhchem}
\hypersetup{colorlinks=true, linkcolor=blue, citecolor=red, urlcolor=blue}

\everymath{\displaystyle}

\newcommand{\0}{\bm{0}}
\newcommand{\GG}{\bm{G}}
\newcommand{\RR}{\bm{R}}
\renewcommand{\aa}{\bm{a}}
\newcommand{\kk}{\bm{k}}
\newcommand{\nn}{\bm{n}}
\newcommand{\pp}{\bm{p}}
\newcommand{\qq}{\bm{q}}
\newcommand{\rr}{\bm{r}}
\newcommand{\uu}{\bm{u}}
\newcommand{\DDelta}{\bm{\Delta}}
\newcommand{\PPi}{\bm{\Pi}}
\newcommand{\ssigma}{\bm{\sigma}}
\newcommand{\nnabla}{\bm{\nabla}}

\newcommand{\mcA}{\mathcal{A}}
\newcommand{\mcB}{\mathcal{B}}
\newcommand{\mcC}{\mathcal{C}}
\newcommand{\mcD}{\mathcal{D}}
\newcommand{\mcI}{\mathcal{I}}
\newcommand{\mcL}{\mathcal{L}}
\newcommand{\mcO}{\mathcal{O}}
\newcommand{\mcT}{\mathcal{T}}
\newcommand{\mcU}{\mathcal{U}}
\newcommand{\mmcA}{\bm{\mathcal{A}}}
\newcommand{\mmcR}{\bm{\mathcal{R}}}

\newcommand{\ffrakA}{\bm{\mathfrak{A}}}

\newcommand{\Id}{\mathbbm{1}}
\newcommand{\LLL}{{\rm LLL}}
\newcommand{\tr}{{\rm tr}}

\newcommand{\Abs}[1]{\left|{#1}\right|}

\newcommand{\bra}[1]{\langle{#1}|}

\newcommand{\ket}[1]{|{#1}\rangle}
\newcommand{\Ket}[1]{\left|{#1}\right\rangle}
\newcommand{\Braket}[2]{\left\langle{#1}\middle|{#2}\right\rangle}

\newcommand{\Braoperket}[3]{\left\langle{#1}\middle|{#2}\middle|{#3}\right\rangle}

\begin{document}

\title{Adiabatic Approximation and Aharonov-Casher Bands in Twisted Homobilayer TMDs}

\author{Jingtian Shi}
\affiliation{Department of Physics, University of Texas at Austin, Austin TX 78712, USA}

\author{Nicol\'as Morales-Dur\'an}
\affiliation{Department of Physics, University of Texas at Austin, Austin TX 78712, USA}

\author{Eslam Khalaf}
\affiliation{Department of Physics, Harvard University, Cambridge, Massachusetts 02138, USA}

\author{A.H. MacDonald}
\affiliation{Department of Physics, University of Texas at Austin, Austin TX 78712, USA}

\date{\today}

\begin{abstract}

Topological flat moir\'e bands with nearly ideal quantum geometry have been identified in homobilayer transition metal dichalcogenide moir\'e superlattices, and are thought to be crucial 
for understanding the fractional Chern insulating states recently observed therein.  
Previous work proposed viewing the system using an adiabatic approximation that replaces the 
position-dependence of the layer spinor with a nonuniform periodic effective magnetic field.
When the local zero-point kinetic energy of this magnetic field cancels identically against that of an effective Zeeman energy,
a Bloch-band version of Aharonov-Casher zero-energy modes, which we refer to as Aharonov-Casher band, emerges leading to ideal quantum geometry. Here, we critically examine the validity of the adiabatic approximation and identify the parameter regimes under which Aharonov-Casher bands emerge. We show that the adiabatic approximation is accurate for a wide range of parameters including those realized in experiments. Furthermore, we show that while the cancellation leading to the emergence of Aharonov-Casher bands is generally not possible beyond the leading Fourier harmonic,
the leading harmonic is the dominant term in the Fourier expansions of the zero-point kinetic energy and Zeeman energy. As a result, the leading harmonic expansion accurately captures the trend of the bandwidth and quantum geometry, though it may fail to quantitatively reproduce more detailed information about the bands such as the Berry curvature distribution.

\end{abstract}

\maketitle

\section{Introduction}

\begin{figure}
    \centering
    \includegraphics[width=0.48\textwidth]{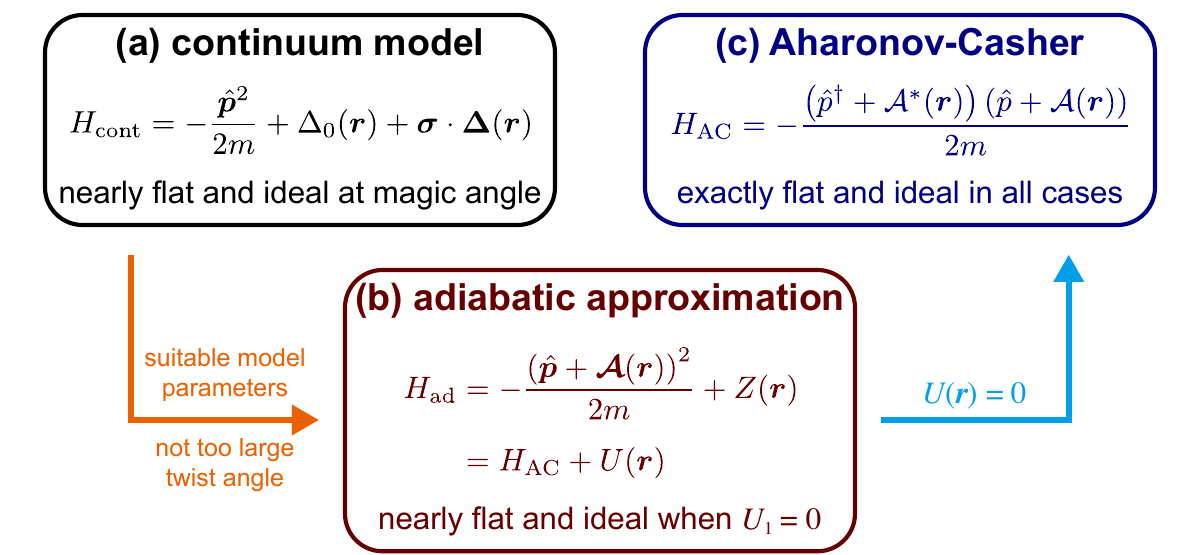}
    \caption{An outline of various approximation regimes we use to understand the ideal flat band behavior of twisted homobilayer TMDs. Here $\ssigma = (\sigma^x, \sigma^y, \sigma^z)$ is the layer-pseudospin Pauli matrix vector, $\hat p = \hat p_x + i\hat p_y = -i\partial_x + \partial_y$ ($\hbar = 1$) and $\mcA(\rr) = \mcA_x(\rr) + i\mcA_y(\rr)$. Details are explained in the text.}
    \label{fig_outline}
\end{figure}

The recent observation of zero-magnetic field fractional Chern insulating (FCI) states \cite{cai2023signatures, zeng2023thermodynamic, park2023observation, xu2023observation, lu2024fractional} has triggered tremendous interest in the 
strongly correlated states of twisted transition metal dichalcogenide (TMD) homobilayer moir\'e superlattices \cite{xu2024maximally, song2024phase, dong2023composite, abouelkomsan2024band, wang2024fractional, yu2024fractional}.
In parallel-stacked homobilayer TMDs with a small twist as we will focus on throughout this paper, the standard continuum model \cite{wu2019topological} of TMD homobilayer moir\'e superlattices predicts a magic twist angle range \footnote{
In this paper we will adopt the 
convention of referring to twist angles at which band width and quantum geometry are both favorable 
for fractional quantum Hall physics as magic angles.  Neither property is ever exactly ideal in realistic models
and the properties are never optimized at precisely the same twist angle.  We will define the magic twist angle as that at which the band width is 
minimized.  Fractional quantum Hall physics is expected over a range of twists surrounding the magic angle whose width is not yet 
experimentally established.
} at which the first valence valley-projected moir\'e band is nearly flat \cite{devakul2021magic, li2021spontaneous, crepel2023anomalous} and topologically nontrivial \cite{wu2019topological, pan2020band, li2021spontaneous} with
nearly ideal quantum geometry \cite{morales-duran2023pressureenhanced}.
These properties are
believed to be key \cite{roy2014band, jackson2015geometric, claassen2015positionmomentum, lee2017band, ledwith2020fractional, ozawa2021relations, mera2021kahler, mera2021engineering, wang2021exact, ledwith2023vortexabilitya, liu2024recent} to FCI states at fractional band fillings.
In a recent Letter \cite{morales-duran2024magic}, some of us have proposed an explanation for 
these properties based on an adiabatic approximation that recognizes the non-trivial topological character of the 
layer-pseudospin field in TMD homobilayers \cite{wu2019topological} and 
assumes that the layer pseudospin is locked to the local direction of the 
model's pseudospin field $\DDelta(\rr)$ (see Fig. \ref{fig_outline} (a)).
A similar approximation has been adopted in the past \cite{ye1999berry,ohgushi2000spin,hamamoto2015quantized,vanhoogdalem2013magnetic} 
to speculate on the possibility of quantum hall effects in thin-film Skyrmion crystals. 
In this approximation, the real-space Berry phase of the layer pseudospin is represented by a position-dependent 
effective magnetic field \cite{yu2020giant, zhai2020theory, li2024contrasting} which has a nonzero average value with one flux quantum per moir\'e unit cell.
The adiabatic approximation is accurate in the small twist angle limit where the moir\'e period is large,
and is shown in this work to maintain accuracy near the magic twist angles 
of homobilayer TMD moir\'e models with experimentally realistic model parameters \cite{devakul2021magic}.

The adiabatic approximation Hamiltonian $H_{\rm ad}$ is that of a 2-dimensional (2D) hole with 
a periodic (dimensionless) effective magnetic field $\mcB(\rr)$ and a periodic potential $Z(\rr)$
that we view as a position dependent effective Zeeman energy (see Fig. \ref{fig_outline} (b)).
Our picture of TMD homobilayers draws on the observation of Aharonov and Casher \cite{aharonov1979ground} that a particle in an arbitrary (not necessarily uniform) magnetic field has a number of zero energy states \cite{estienne2023ideal, crepel2024chirala, parhizkar2024generic}
-- one for each quantum of magnetic flux --
when the zero point kinetic energy $\mcB(\rr)/2m$ cancels locally against the Zeeman energy $Z(\rr)$.
In the case considered by Aharonov and Casher, this cancellation is a property of a spin-$1/2$ electron
with spin g-factor $g=2$ where 
$Z(\rr)$ is simply the true Zeeman energy produced by the magnetic field.
In the case of interest here, the effective magnetic field does not 
couple to spin and the cancelling $Z(\rr)$ must have a different origin.  When cancellation does occur (see Fig. \ref{fig_outline} (c)),
a special case of the situation considered by Aharonov and Casher is presented in which the 
position-dependent magnetic field has crystal 
translational symmetry and the zero energy states therefore form a (quasi-) Bloch band \cite{dubrovin1980ground}, which we refer to as the Aharonov-Casher (AC) band.
Since each AC zero energy
state has the form of a holomorphic function of $z=x+iy$ times a common factor, the AC band is vortexable \cite{ledwith2023vortexabilitya}
and has ideal quantum geometry, and therefore \cite{claassen2015positionmomentum, mera2021engineering, wang2021exact} wave functions similar to those of Landau levels.

\begin{figure}
	\centering
	\includegraphics[width=0.48\textwidth]{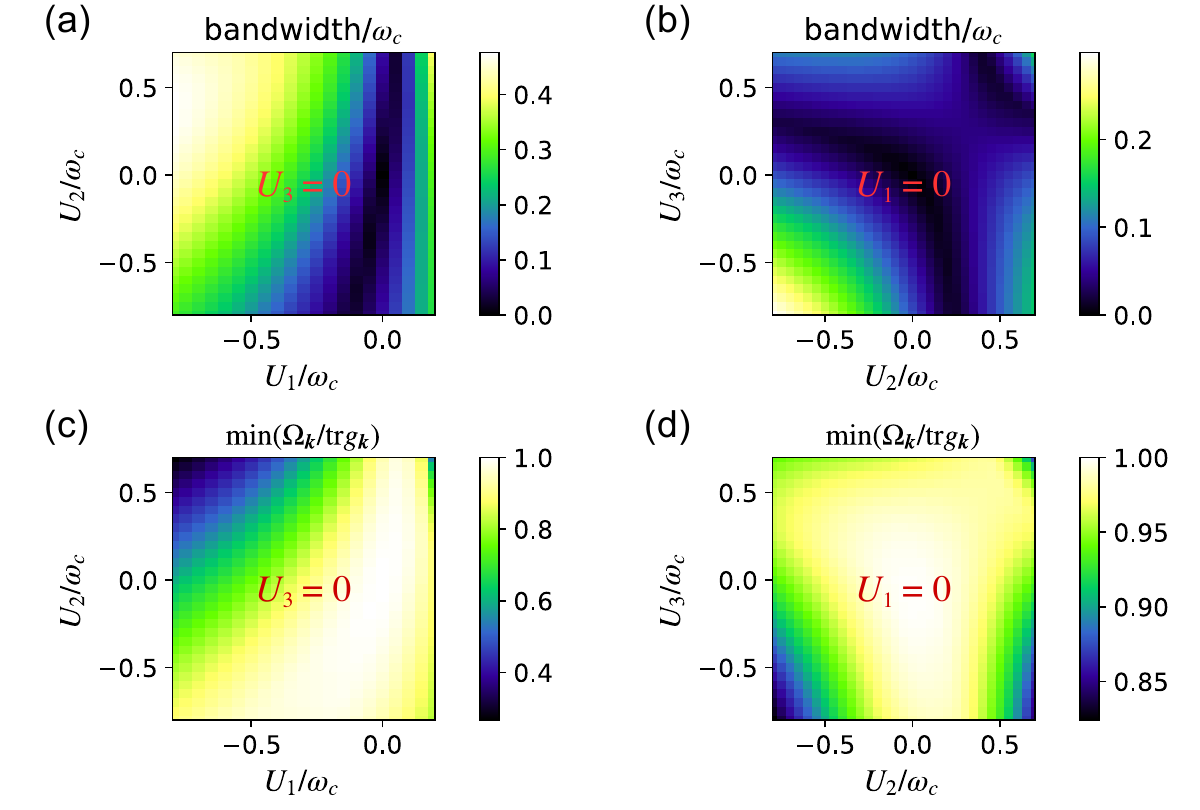}
	\caption{(a)-(b) Dependence of the bandwidth of the first moir\'e miniband of $H_{\rm ad}$ on $U(\rr)$ characterized by its Fourier coefficients $U_i$, where $i$ labels a shell of reciprocal lattice vectors.
    We have taken the typical \ce{WSe_2} model parameter from Ref. \onlinecite{devakul2021magic} and assumed $U_i=0$ for all $i>3$ and
    (a) $U_3 = 0$; (b) $U_1 = 0$. (c)-(d) The quantum geometry idealness of the first band under the same parameter settings. Here the idealness is characterized by the minimum over the entire
moir\'e Brillouin-zone (mBZ) of the ratio of the Berry curvature $\Omega_{\kk}$ to the trace of the Fubini-Study metric $g_{\kk}$.
	This ratio is always smaller than or equal to 1 \cite{roy2014band}, and is identically equal to 1 for an ideal band.
	The plot ranges are selected within the region of Chern number $1$.}
	\label{fig_theme1}
\end{figure}

In TMD homobilayer moir\'es, the local Zeeman energy $Z(\rr)$ never cancels the 
zero-point energy perfectly.  We refer to the difference as the residual potential $U(\rr)$ (see Fig. \ref{fig_outline} (b)).
It follows from the emergent honeycomb lattice symmetry of $H_{\rm ad}$ that the residual potential is characterized by Fourier components $U_i$ that 
are real and constant within each shell of the reciprocal lattice, labeled by index $i$.
Although a perfect flat ideal band is realized only when all $U_i=0$,
we show in this paper that $U_1=0$ is usually sufficient to produce narrow bands with nearly ideal quantum geometry.
This property is illustrated in Fig. \ref{fig_theme1}.  The Landau-level like regime can therefore be reached by 
varying a single tuning parameter, for example twist angle.  
Interestingly though, we find that the higher Fourier components of $U(\rr)$ can significantly alter the wave 
function of the AC band while keeping the quantum geometry nearly ideal.

This paper is organized as follows. In Secs. \ref{subsec_cont} and \ref{subsec_ad} we specify the continuum model and adiabatic approximation Hamiltonians
of twisted homobilayer TMDs, respectively, and comment on their properties and relevant model parameters appropriate for tungsten diselenide (\ce{WSe_2}) and molybdenum ditelluride (\ce{MoTe_2}). 
In Sec. \ref{subsec_AC} we describe the properties of the AC band.
In Sec. \ref{subsec_LL} we introduce the numerical method we use to diagonalize $H_{\rm ad}$ using a Landau level basis. 
In Sec. \ref{sec_comp_ad_cont} we 
demonstrate the validity of adiabatic approximation in near-magic-angle \ce{WSe_2} by comparing the adiabatic approximation and continuum model results for the band structure, bandwidth, quantum geometry, Berry curvature distribution and full-band charge density distribution.
In Sec. \ref{sec_comp_AC_ad} we discuss
the influence of the residual potential $U(\rr)$ on the full-band charge density distribution and Berry curvature, taking the \ce{WSe_2} parameters \cite{devakul2021magic} under the adiabatic approximation magic angle as an example.
In Sec. \ref{subsec_otherbands} we show that the adiabatic approximation accurately
reproduces the topological phases of higher energy bands as a function of twist angle, 
and point to a transition between Landau-level-like and Haldane-model-like \cite{haldane1988model} band structures, 
which has been previously identified \cite{wu2019topological, wang2023staggered, crepel2023anomalous, liu2024gatetunable} in the continuum model. 
In Sec. \ref{sec_conclusion} we conclude and discuss some possible future developments.

\section{Models}

\subsection{Continuum Model}
\label{subsec_cont}

The continuum model Hamiltonian \cite{wu2019topological} of the valence bands of TMD homobilayers has the form of a hole with effective mass $m$
in the presence of a spatially periodic scalar potential $\Delta_0(\rr)$ and a 
spatially periodic vector field $\DDelta(\rr)$ that couples to the layer pseudospin (see Fig. \ref{fig_outline} (a)).
Here
\begin{subequations}
	\label{eq_contDelta}
    \begin{gather}
        \Delta_0(\rr) \pm \Delta_z(\rr) = 2V \sum_{j=0}^2 \cos \left( \GG_{2j}\cdot\rr \mp\psi \right), \\\addlinespace
        \Delta_x(\rr) \pm i\Delta_y(\rr) = w\sum_{j=0}^2 e^{\pm i\qq_j\cdot\rr},
    \end{gather}
\end{subequations}
where $\qq_j = q ( \sin(2j\pi/3), \, -\cos(2j\pi/3) )$ and $\GG_j = \sqrt{3}q ( \cos(j\pi/3), \, \sin(j\pi/3) )$ are the interlayer and intralayer transfer momenta due to the moir\'e potential, $q = 4\pi/3a_M$ is the radius of the mBZ, $a_M = a/2\sin(\theta/2)$ is the moir\'e lattice constant of the superlattice at twist angle $\theta$ and $a$ is the lattice constant of the single-layer material.
Note that we have chosen a $C_3$ symmetric gauge \cite{morales-duran2024magic} that is different than in previous literature \cite{wu2019topological}.

The material-specific model parameters $w$, $V$ and $\psi$ respectively characterize
the strength of interlayer tunneling, the strength of the potential energy in each layer and a phase angle that captures the positions at which the potential has its extrema.
We note that a sign flip in the basis on either layer leads to $w \rightarrow -w$, and that since the two valleys are related by time-reversal symmetry $\mcT$, a combined operation of in-plane 2-fold rotation $C_2$ and time reversal $\mcT$ ($C_2\mcT$) does not change the physics while it brings $\Delta(\rr)$ to $\Delta^*(-\rr)$, which results in an overall change in model parameter $\psi \rightarrow -\psi$. In addition, the substitution $(V, \psi) \rightarrow (-V, \psi + 180^\circ)$ leaves the model Hamiltonian unchanged. Hereby, the parameter space is folded into the regime $w > 0$, $V > 0$ and $0 < \psi < 180^\circ$.
A compilation of \textit{ab initio}-based parameter values for \ce{WSe_2} and \ce{MoTe_2} from previous work, after the symmetry folding described above, is provided in Table \ref{table_app_modelparameters} in the appendix. For simulations of \ce{WSe_2} and \ce{MoTe_2}, we respectively take the values from Refs. \onlinecite{devakul2021magic}
$(a, m, w, V, \psi) = (3.317\text{\AA}, 0.43m_e, 18{\rm meV}, 9{\rm meV}, 128^\circ)$ ($V/w = 0.5$)
and \onlinecite{wang2024fractional}
$(a, m, w, V, \psi) = (3.52\text{\AA}, 0.6m_e, 23.8{\rm meV}, 20.8{\rm meV}, 107.7^\circ)$ ($V/w = 0.87$), assuming them to be twist angle independent, though they certainly do
depend on the twist angle when moir\'e relaxation strains are accounted for.  Here $m_e$ is the free electron mass.
Some work has also taken into account higher harmonics of moir\'e tunneling and potential terms \cite{zhang2024polarizationdrivena, mao2024transfer, jia2024moire} and strain-induced gauge field correction to kinetic energy terms \cite{mao2024transfer}.  Although these can improve models for specific materials, we choose to ignore them here to focus 
on the small number of parameters that have the greatest importance.

\subsection{Adiabatic Approximation for Twisted Homobilayer TMDs}
\label{subsec_ad}

The adiabatic approximation Hamiltonian \cite{morales-duran2024magic} $H_{\rm ad}$ is obtained from a layer-pseudospin $\rm U(2)$ gauge transformation of the continuum model Hamiltonian $H_{\rm cont}$ that locally rotates the vector field $\DDelta(\rr)$ to the out-of-plane $+z$ direction:
\begin{equation}
	\mcU^\dag(\rr) \left[ \ssigma\cdot\DDelta(\rr) \right] \mcU(\rr) = \Abs{\DDelta(\rr)} \sigma^z,
	\label{eq_U2gauge}
\end{equation}
leading to an emergent non-Abelian connection $\ffrakA(\rr) = -i\mcU^\dag(\rr) \nnabla\mcU(\rr)$, followed by a simple projection into the (rotated) spin-up sector. The approximation is valid at small twist angles, where the smooth spatial variation of the vector field ensures good spatial adiabaticity. The detailed derivation is presented in Appendix \ref{sec_app_ad}.
$H_{\rm ad}$ can be written in the following equivalent forms:
\begin{subequations} \label{eq_ad_Ham} \begin{align}
	H_{\rm ad} &= -\frac{\left( \hat\pp + \mmcA(\rr) \right)^2}{2m} + Z(\rr) \label{eq_ad_Ham0} \\\addlinespace
	&= -\frac{\left( \hat p^\dag + \mcA^*(\rr) \right) \left( \hat p + \mcA(\rr) \right)}{2m} + U(\rr) \label{eq_ad_Ham1} \\\addlinespace
	&= -\frac{\hat\Pi_0^\dag \hat\Pi_0}{2m} - \frac{\hat\Pi_0^\dag \mcA'(\rr) + \mcA'^*(\rr) \hat\Pi_0}{2m} + U'(\rr), \label{eq_ad_Ham2}
\end{align} \end{subequations}
where $\mmcA(\rr)= \ffrakA_{++}(\rr)$ is the vector potential of the emergent effective magnetic field $\mcB = \nnabla \times \mmcA$;
$\hat p = \hat p_x + i\hat p_y = -2i\partial_{z^*}$, $\hat p^\dag = \hat p_x - i\hat p_y = -2i\partial_z$, $z = x + iy$, $\mcA(\rr) = \mcA_x(\rr) + i\mcA_y(\rr) = \mcA_0(\rr) + \mcA'(\rr)$, $\mcA_0(\rr) = i\mcB_0z/2$ is the vector potential of the spatial average $\mcB_0 = -2\pi/A_M$ of the magnetic field, $\hat\Pi_0 = \hat p + \mcA_0(\rr)$, $A_M = \sqrt{3} a_M^2/2$ is the area of the moir\'e unit cell, $a_M = a/2\sin(\theta/2) \approx a/\theta$ is the moir\'e period, $a$ is the lattice constant of the TMD material and $\theta$ is the twist angle.
The potential terms in Eqs. (\ref{eq_ad_Ham}) have the form
\begin{subequations} \begin{gather}
	Z(\rr) = \Delta_+(\rr) -\frac{\mcD(\rr)}{2m} = \Delta_+(\rr) - \omega_c d(\rr), \label{eq_Z(r)} \\\addlinespace
	U(\rr) = Z(\rr) + \frac{\mcB(\rr)}{2m} = \Delta_+(\rr) - \omega_c \xi(\rr),
	\label{eq_ideal_pseudopotential} \\\addlinespace
	U'(\rr) = U(\rr) - \frac{\Abs{\mcA'(\rr)}^2}{2m} = \Delta_+(\rr) - \omega_c \xi'(\rr), \label{eq_LL_U'}
\end{gather} \end{subequations}
where $\Delta_+(\rr) = \Delta_0(\rr) + |\DDelta(\rr)|$ is the local energy of the layer pseudospin that is aligned with the moir\'e skyrmion field, 
$\omega_c = |\mcB_0|/m \propto \theta^2 \propto A_M^{-1}$ is the spacing between the Landau levels defined by $\mcB_0$, and $\mcD(\rr)= |\ffrakA_{+-}(\rr)|^2$
is a kinetic potential that arises from off-diagonal elements of $\ffrakA(\rr)$ and turns out to be exactly the trace of the real-space quantum metric of the layer-pseudospin skyrmion field texture, as shown in Appendix \ref{sec_app_ad}.
\footnote{For a summary of notation difference between this manuscript and Ref. \onlinecite{morales-duran2024magic}: under our convention that $\hbar=1$, $\mmcA$, $\mcB$, $\mcD$, $\Delta_+$ and $m$ in this paper correspond respectively to $e\tilde{\bm A}$, $eB_{\rm eff}^z$, $2m^*D$, $\tilde\Delta$ and $m^*$ in Ref. \onlinecite{morales-duran2024magic}.}
The amplitude of $\Delta_+(\rr)$ is independent of 
twist angle for a given $(w,V,\psi)$ as are those of the scaled fields:
\begin{subequations}
	\begin{align}
		d(\rr) & = \frac{A_M\mcD(\rr)}{4\pi}, \\\addlinespace
		\xi(\rr) & = \frac{A_M}{4\pi} \left[ \mcD(\rr) - \mcB(\rr) \right],  \\\addlinespace
		\xi'(\rr) &= \frac{A_M}{4\pi} \left[ \mcD(\rr) + \Abs{\mcA'(\rr)}^2 - \mcB(\rr) \right].
	\end{align}
\end{subequations}

\begin{figure}
	\centering
	\includegraphics[width=0.48\textwidth]{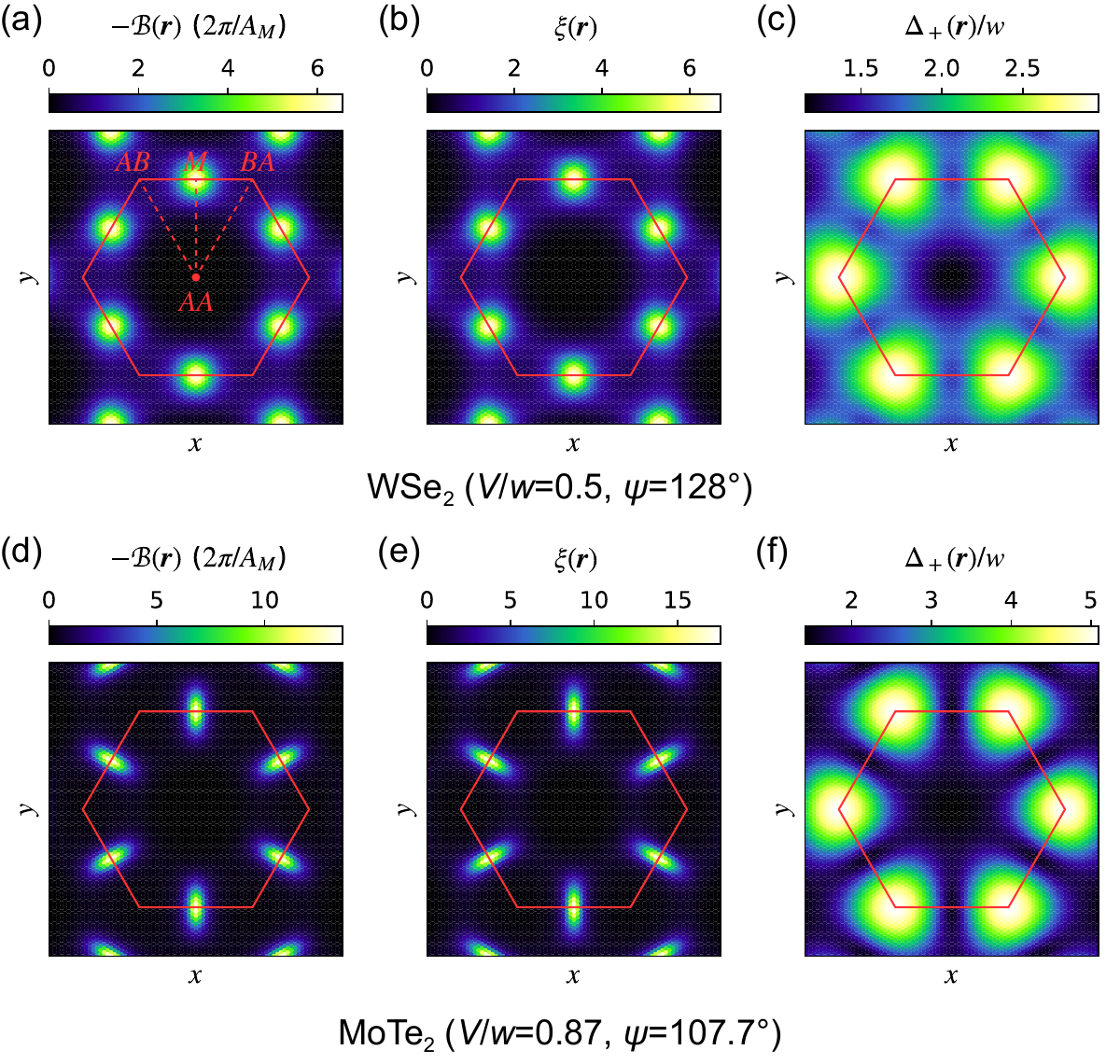}
	\caption{(a)-(c) Plots of (a) the dimensionless effective magnetic field $-\mcB(\rr) A_M/2\pi$, (b) the dimensionless quantity $\xi(\rr)$ and (c) the effective Zeeman energy $\Delta_+(\rr)$ (in units of $w$) in the adiabatic approximation calculated using \ce{WSe_2} parameter estimates $(w, V, \psi) = (18{\rm meV}, 9{\rm meV}, 128^\circ)$ \cite{devakul2021magic}. (d)-(f) The same plots for \ce{MoTe_2} with model parameters $(w, V, \psi) = (23.8{\rm meV}, 20.8{\rm meV}, 107.7^\circ)$ \cite{wang2024fractional}. In all panels, the hexagon is the Wigner-Seitz moir\'e unit cell. The high-symmetry positions $AA$, $AB$ (chalcogen on metal for \ce{WSe_2}, metal on chalcogen for \ce{MoTe_2}), $BA$, and Wigner-Seitz edge mid-point $M$ are marked in (a).}
	\label{fig_realspace}
\end{figure}

For the \ce{WSe_2}\cite{devakul2021magic} and \ce{MoTe_2}\cite{wang2024fractional} parameter sets we have taken, we respectively have $\omega_c \approx 3.56{\rm meV}\cdot(\theta[^\circ])^2$ and $\omega_c \approx 2.27{\rm meV}\cdot(\theta[^\circ])^2$.
The spatial structure of some real-space fields employed in the adiabatic approximation that are calculated from these %\ce{WSe_2} and \ce{MoTe_2} 
parameter sets is illustrated
in Fig. \ref{fig_realspace}. The two materials have similar shapes of the functions $|\mcB(\rr)|$, $\xi(\rr)$ and $\Delta_+(\rr)$, and these 
are also similar to the results \cite{morales-duran2024magic} obtained from non-relaxed \ce{MoTe_2} parameters \cite{wu2019topological}. We note that the fields are generally more sharply peaked in \ce{MoTe_2} than in \ce{WSe_2}.

\subsection{The Aharonov-Casher Band}
\label{subsec_AC}

When $U(\rr)$ in Eq. (\ref{eq_ad_Ham1}) vanishes identically, the analysis of Aharonov and Casher \cite{aharonov1979ground, dubrovin1980ground}
implies that the Hamiltonian has a perfect zero-energy eigenspace with wave functions of the form
\begin{equation}
	\psi^{\rm AC}(\rr) = f(z) e^{\alpha(\rr)} = f(z) e^{-\frac{\rr^2}{4\ell^2} + \chi(\rr)} = e^{\chi(\rr)} \psi^{\LLL}(\rr),
	\label{eq_AC_wavefunction}
\end{equation}
where $f$ is arbitrary holomorphic function, similar to those in lowest Landau level (LLL) wave functions \cite{girvin1984formalism}, and $\partial_{z^*}\alpha = i\mcA/2$. Out of the remaining $\rm U(1)$ gauge freedom from the local diagonalization in Eq. (\ref{eq_U2gauge}), we choose a symmetric gauge with $\nnabla\cdot\mmcA = 0$ such that the real periodic function $\chi(\rr)$ can be found by solving $\nnabla^2\chi(\rr) = \mcB'(\rr) = \mcB(\rr) - \mcB_0$ (or $\nabla^2\alpha(\rr) = \mcB(\rr)$), where $\mcB'(\rr)$ is a periodic function with zero average.

Since $\chi(\rr)$ is periodic, there exists a complete manifold of magnetic quasi-Bloch wave functions in the zero-energy eigenspace, $\psi_{\kk}^{\rm AC}(\rr) = e^{\chi(\rr)} \psi_{\kk}^{\LLL}(\rr)$, that is reminiscent of the quasi-Bloch construction of LLL wave functions \cite{ferrari1990twodimensional, ferrari1995wannier, haldane2018modularinvariant, wang2019lattice, wang2021exact}:
\begin{subequations}
	\label{eq_LLLquasiBloch}
	\begin{equation}
		\psi_{\kk}^{\LLL}(\rr) = e^{\frac{i}{2} \kk\cdot\rr} \psi_{\0}^{\LLL} (\rr + \ell^2\nn_z\times\kk),
		\label{eq_psikLLL}
	\end{equation}
	\begin{equation}
		\psi_{\0}^{\LLL}(\rr) = \pi^{\frac{1}{4}} \sqrt{\frac{2\ell}{|a_1|}} e^{\frac{1}{4\ell^2} \left( \frac{a_1^*z^2}{a_1} - \rr^2 \right)}
		\vartheta_1 \left( \frac{z}{a_1} \middle| \frac{a_2}{a_1} \right),
		\label{eq_psi0LLL}
	\end{equation}
	\begin{equation}
		\vartheta_1 (\zeta|\tau) \equiv \sum_{n=-\infty}^\infty
		e^{i\pi \left[ (2n+1) \left( \zeta + \frac{1}{2} \right) + \left( n + \frac{1}{2} \right)^2 \tau \right]},
	\end{equation}
\end{subequations}
where $\nn_z$ is the out-of-plane unit vector in $+z$ direction, $\ell = 1/\sqrt{|\mcB_0|} = \sqrt{A_M/2\pi}$ is the magnetic length, $a_j = a_{jx} + ia_{jy}$ for $j = 1,2$ with $\aa_1$, $\aa_2$ the primitive basis vectors of the moir\'e superlattice, and $\vartheta_1$ is the (auxiliary) Jacobi theta function. Here we have chosen the normalization convention that
\begin{equation}
	\Braket{\psi_{\kk}^{\LLL}}{\psi_{\kk}^{\LLL}} = \frac{1}{A} \int d^2\rr \Abs{\psi_{\kk}^{\LLL}(\rr)}^2 = 1
\end{equation}
where $A$ is the macroscopic system area. The quasi-Bloch property of $\psi_{\kk}^{\rm AC}$ (and also $\psi_{\kk}^{\LLL}$) is captured by that
it is an eigenstate of all magnetic translational operators $\hat T_{\RR}$ on the moir\'e Bravais lattice:
\begin{equation}
	\hat T_{\RR} \psi_{\kk}^{\rm AC}(\rr) = e^{\frac{i\RR\times\rr}{2\ell^2}} \psi_{\kk}^{\rm AC} (\rr-\RR)
	= \eta_{\RR} e^{-i\kk\cdot\RR} \psi_{\kk}^{\rm AC}(\rr),
\end{equation}
where $\eta_{\RR} = (-1)^{\mu + \nu + \mu\nu}$ is the parity of lattice vector $\RR = \mu\aa_1 + \nu\aa_2$ in the Bravais lattice, \textit{i.e.} 1 if $\RR/2$ is in the Bravais lattice and $-1$ otherwise.

The quasi-Bloch manifold forms the AC band, which has Chern number $C = 1$ and perfectly ideal quantum geometry following from its holomorphic algebra inherited from the LLL. According to well-established theories of ideal bands\cite{wang2021exact}, 
the AC band Berry curvature is given by
\begin{subequations} \label{eq_Berrycurv_AC} \begin{gather}
	\Omega_{\kk}^{\rm AC} = \frac{2\pi}{A_{\rm BZ}} + \frac{1}{2} \nnabla_{\kk}^2 \ln \Braket{\psi_{\kk}^{\rm AC}}{\psi_{\kk}^{\rm AC}},
	\label{eq_Berrycurv_ACa}
	\\\addlinespace
	\begin{aligned}
		\Braket{\psi_{\kk}^{\rm AC}}{\psi_{\kk}^{\rm AC}} &= \Braoperket{\psi_{\kk}^{\LLL}}{e^{2\chi(\rr)}}{\psi_{\kk}^{\LLL}} \\\addlinespace
		&= \sum_{\GG} \overline\eta_{\GG} \lambda_{\GG} \Phi_{\GG} e^{i\ell^2\GG\times\kk},
	\end{aligned} \label{eq_Berrycurv_ACb}
\end{gather} \end{subequations}
where $A_{\rm BZ} = 4\pi^2/A_M = 2\pi/\ell^2$ is the area of mBZ, $\overline\eta_{\GG}$ is the parity of $\GG$ in the reciprocal lattice (\textit{i.e.} $1$ if $\GG/2$ is a reciprocal lattice vector and $-1$ otherwise), the magnetic form factor $\lambda_{\GG} = e^{-\ell^2\GG^2/4}$, and $\Phi_{\GG}$ is the Fourier component of $e^{2\chi(\rr)}$.
Note that this Berry curvature is independent of twist angle up to a factor of $A_{BZ}^{-1}$, which is a feature specific to the AC limit.

\subsection{Landau Level Representation}
\label{subsec_LL}

The perfect AC limit is only reached when $U(\rr) = 0$, which requires an identical cancellation between $\Delta_{+}(\rr)$ and $\xi(\rr)$, as seen from Eq. (\ref{eq_ideal_pseudopotential}). Importantly we see in Fig. \ref{fig_realspace} that these two functions are both peaked near the Wigner-Seitz cell boundary but at different positions.  Their relative scale is dictated by twist angle, but perfect 
cancellation is impossible because of this shape mismatch. (We have identified a region in the continuum model parameter 
space in which $U(\rr)$ is small so that $H_{\rm AC}$ is a good approximation to $H_{\rm ad}$, which is discussed in Appendix \ref{sec_app_res_pot}. Unfortunately, that is also a region where the adiabatic approximation fails.)
With the mismatch, the band structure of $H_{\rm ad}$ must be calculated numerically.

We express $H_{\rm ad}$ in the representation of the 
Landau levels defined by $\mcB_0$ and converge its spectrum with respect to the Landau level cutoff.
Because there is one quantum of flux in the unit cells defined by the 
periodicities of $\mcA'(\rr)$ and $U'(\rr)$, we can use a convenient Landau level quasi-Bloch basis, with the lowest Landau-level wave function $\psi_{\kk}^{\LLL}(\rr)$ specified by Eqs. (\ref{eq_LLLquasiBloch}) and
higher Landau level states constructed by using Landau level raising operators
\begin{equation}
	\Ket{\psi_{\kk}^{n\rm LL}} = \frac{\left( a_0^\dag \right)^n}{\sqrt{n!}} \Ket{\psi_{\kk}^{\LLL}}.
	\label{eq_psik_nLL_raisedbya0}
\end{equation}
Like $\ket{\psi_{\kk}^{\LLL}}$, the states $\ket{\psi_{\kk}^{n\rm LL}}$ are also eigenstates of the magnetic translation operator with 
quasimomentum $\kk$, so that states with different $\kk$'s in the mBZ are decoupled. 
In Eq. (\ref{eq_psik_nLL_raisedbya0}), $a_0^\dag = \ell \hat\Pi_0^\dag/\sqrt{2}$ is the Landau level raising operator
and $n\rm LL$ stands for the $n$th Landau level. 
In this representation the first term in Eq. (\ref{eq_ad_Ham2}) gives the diagonal Landau level kinetic
energy contribution, and the sum of the second 
and third terms, denoted $H'$, is
\begin{widetext}
	\begin{equation}
		\Braoperket{\psi_{\kk}^{n\rm LL}}{H'}{\psi_{\kk}^{n'\rm LL}} = \frac{(-1)^n}{\sqrt{n!n'!}} \sum_{\GG} \overline\eta_{\GG} \lambda_{\GG} \left(
			\mcL_{nn'}(\Gamma^*, \Gamma) \, U_{\GG}' + \left(
				\frac{n\mcL_{(n-1)n'}(\Gamma^*, \Gamma)}{\Gamma^*} + \frac{n'\mcL_{n(n'-1)}(\Gamma^*, \Gamma)}{\Gamma}
			\right) \frac{\mcB_{\GG}'}{2m}
		\right) e^{i\ell^2\GG\times\kk}.
		\label{eq_LLformulation_Hamelement}
	\end{equation}
\end{widetext}
(See Appendix \ref{sec_app_LL} for a detailed derivation.)  
In Eq. (\ref{eq_LLformulation_Hamelement}),
$\Gamma = \ell (G_x + iG_y) / \sqrt{2}$, the bivariate polynomial $\mcL_{nn'}(x, y) = e^{xy} \partial_x^n \partial_y^{n'} e^{-xy}$ is 
related to the generalized Laguerre polynomial $L_n^{(\alpha)}$ by $\mcL_{nn'}(x, y) = (-1)^{n'} n! x^{n'-n} L_n^{(n'-n)}(xy)$, and $U_{\GG}'$ and $\mcB_{\GG}'$ are respectively the Fourier components of $U'(\rr)$ and $\mcB'(\rr)$.
Because both $U'(\rr)$ and $\mcB'(\rr)$ have honeycomb lattice symmetry, their Fourier expansions are 
specified by one real number for each shell of reciprocal lattice vectors.
The exponentially decaying nature of $\lambda_{\GG}$ converges 
the reciprocal lattice sum.  Whereas the first shell is usually sufficent to get 
relatively accurate results \cite{morales-duran2024magic} for $n=n'=0$, higher shells
become important at larger $n$ and/or $n'$ because of the $\mcL_{nn'}(\Gamma^*, \Gamma)$ factors. 
Formulas for Berry curvature and other components of the quantum geometry tensor 
under this Landau level basis,
which will figure importantly in our analysis,
are derived in Appendix \ref{sec_app_Berry_LLbasis}.

\section{Comparison Between Adiabatic and Continuum Model Band Properties}
\label{sec_comp_ad_cont}

We have so far shown that in the adiabatic approximation
the physics of TMD homobilayer moir\'es is mapped to a problem
of scalar electrons in a periodic potential and a periodic magnetic field,
and argued that the adiabatic approximation is accurate in the small twist angle limit.
In this section, we will show by comparing the explicit numerical results that 
for a realistic set of parameters typical of \ce{WSe_2} \cite{devakul2021magic} the adiabatic approximation 
accurately captures a wide range of moir\'e band properties over a large 
range of twist angles up to the magic angle. 

\begin{figure}
	\centering
	\includegraphics[width=0.48\textwidth]{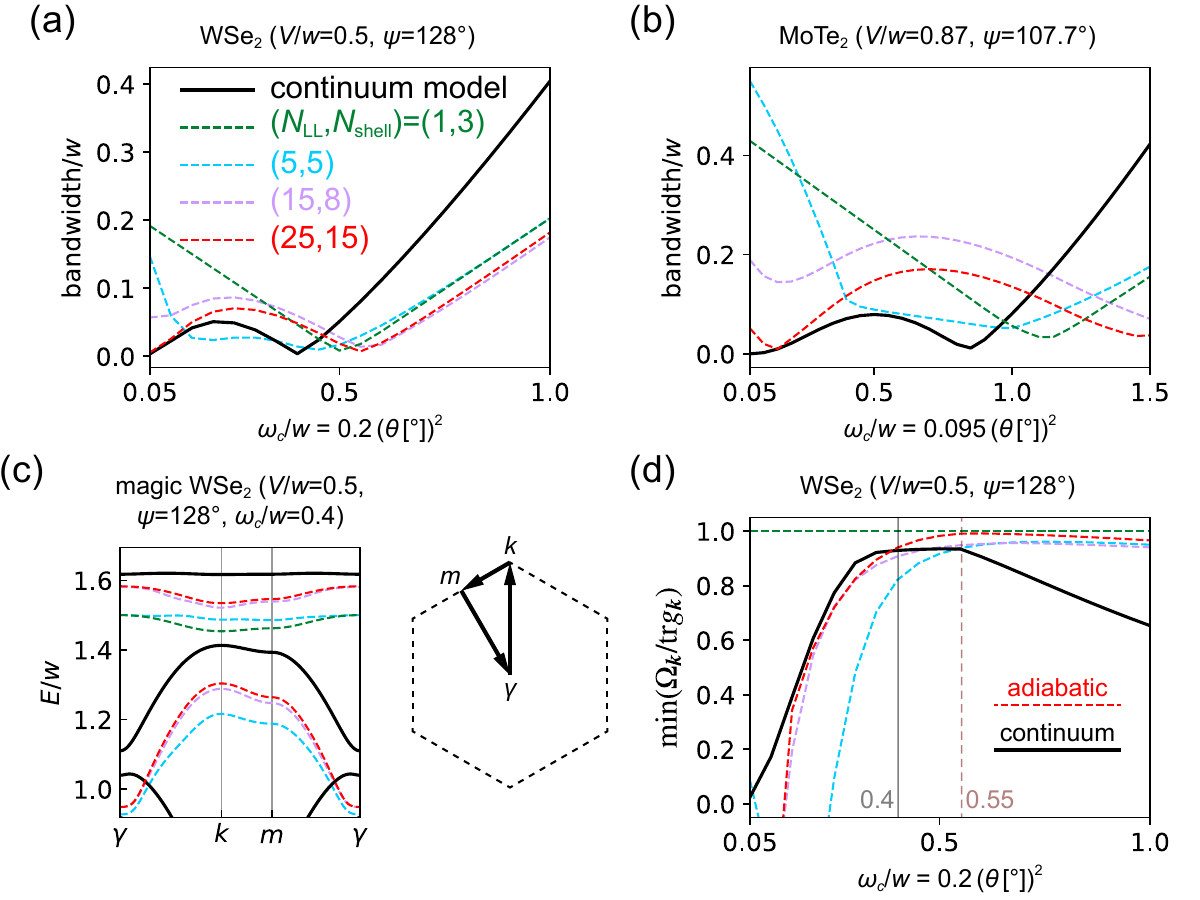}
	\caption{(a)-(b) The width of the first moir\'e valence band of (a) twisted bilayer \ce{WSe_2} and (b) twisted bilayer \ce{MoTe_2} {\it vs.} twist angle, predicted by the continuum model and the adiabatic approximation.  For the adiabatic approximation we use a Landau level representation discussed in Sec. \ref{subsec_LL}.  The plots in (a)-(b) illustrate the 
 dependence on Landau level $N_{\rm LL}$ cutoff.  For each $N_{\rm LL}$ we have indicated the number of momentum space shells 
$N_{\rm shell}$ that we have retained in the Fourier expansions of the potentials and magnetic fields to achieve convergence.
	The color code is defined in (a). 
	(c) Comparison of the continuum model and adiabatic approximation moir\'e band structures of twisted \ce{WSe_2} at 
 $\omega_c/w = 0.4$, plotted along the path shown in the right panel.
	(d) The minimum of the trace condition ratio $\Omega_{\kk}/{\rm tr}g_{\kk}$ over the mBZ as a function of twist angle of \ce{WSe_2}.
	In (d) we relabel the $(25, 15)$ curve as ``adiabatic'' since we have assumed it to be the exact adiabatic solution.
	The solid and dashed vertical lines mark the minimum-bandwidth magic angles in the continuum model and the adiabatic approximation, respectively.}
	\label{fig_comparisons}
\end{figure}

Theoretical studies of homobilayer TMD moir\'es show that the bandwidth narrows and the quantum geometry 
becomes nearly ideal over narrow ranges of twist angle.  (The correspondence between these 
narrow ranges and experimental FCI observations is suspected but not yet established.) Fig. \ref{fig_comparisons}
compares the moir\'e band properties calculated from the continuum model with those calculated in the adiabatic 
approximation using the average field Landau level representation.  Results are 
presented for a variety of different Landau level truncations ($N_{\rm LL}$).
For each $N_{\rm LL}$, the number of shells ($N_{\rm shell}$) retained in the Fourier expansions of $\mcB(\rr)$ and
$U'(\rr)$ is indicated.  Figs. \ref{fig_comparisons} (a) and (b) plot the bandwidths of the first moir\'e valence bands of \ce{WSe_2} and 
\ce{MoTe_2} {\it vs.} the measure of twist angle, 
characterized here by the average-field Landau level energy separation $\omega_c$. Bandwidths  
have an overall trend of increasing with $\omega_c$ as the moir\'e periods get shorter and 
energy scales increase.  In both continuum and adiabatic cases, though, there is a
sharp local minimum that interrupts the increasing trend, which is associated
with the magic angle behavior.
In Fig. \ref{fig_comparisons} (a) we see that the magic angle of twisted bilayer \ce{WSe_2}
defined by minimum bandwidth in the adiabatic approximation ($\omega_{c0}^{\rm ad}/w \approx 0.55$, corresponding to 
$\theta_0^{\rm ad} \approx 1.67^\circ$) is somewhat larger than, but still similar to, 
the continuum model result ($\omega_{c0}^{\rm cont}/w \approx 0.4$, corresponding to 
$\theta_0^{\rm cont} \approx 1.42^\circ$).

Far above the magic angle the adiabatic approximation becomes inaccurate as expected.
We find that the adiabatic approximation generally underestimates the bandwidth. 
Within the adiabatic approximation, more Landau levels are required for convergence at 
smaller twist angles.  The single-Landau level projection ($N_{\rm LL} = 1$) approximation
fails to capture the decrease in bandwidth as the twist angle approaches zero, 
but including more Landau levels reproduces this feature correctly.
For \ce{WSe_2} we find that the adiabatic approximation bandwidth near the 
magic angle converges after including only a small number of Landau levels,
although, as we see in Fig. \ref{fig_comparisons} (c), more levels are needed to converge absolute energies.

We also show results for the bandwidth of the \ce{MoTe_2} model, which has a larger value of $V/w$. 
As Fig. \ref{fig_comparisons} (b) indicates, not only is the adiabatic approximation less 
accurate than in the \ce{WSe_2} case, but the convergence with respect to $N_{\rm LL}$ is slower. 
We argue that the accuracy of adiabatic approximation generally persists to larger twist angles when both $V/w$ and $\psi$ are at intermediate values.
The influence of $V/w$ can be seen by tracking the variation of the layer-pseudospin vector field $\DDelta(\rr)$ along an edge of the real-space Wigner-Seitz cell, which can be obtained from Eq. (\ref{eq_contDelta}):
\begin{subequations} \begin{gather}
	\DDelta(\rr_{AB}) = \left( 0,\ 0,\ -3\sqrt{3}V\sin\psi \right), \\\addlinespace
	\DDelta(\rr_M) = \left( \frac{w}{2},\ \frac{\sqrt{3}w}{2},\ 0 \right), \\\addlinespace
	\DDelta(\rr_{BA}) = \left( 0,\ 0,\ 3\sqrt{3}V\sin\psi \right).
\end{gather} \end{subequations}
Here $\rr_{AB}$, $\rr_{BA}$ and $\rr_{M}$ are respectively the high-symmetry positions $AB$, $M$ and $BA$ labeled in Fig. \ref{fig_realspace} (a). Too large or too small $V/w$ values would cause the direction of $\DDelta(\rr)$ to vary abruptly with position near the real-space $M$ point when 
$V/w$ is large or near the $AB$ and $BA$ points when $V/w$ is small.  In either case adiabaticity applies only at small twist angles 
and sharp peaks appear in the real-space fields $-\mcB(\rr)$ and $\xi(\rr)$ (illustrated in Figs. \ref{fig_realspace} (d) and (e) with the \ce{MoTe_2} parameter set) that mix up more Landau levels. On the other hand, because $\Delta_0(\rr) \propto \cos\psi$, if $\psi$ is close to $0^\circ$ or $180^\circ$, the spatial variation in the scalar field $\Delta_0(\rr)$ will be strong enough to destroy the indirect gap between the (rotated) pseudospin-up and down branches $\Delta_\pm(\rr) = \Delta_0(\rr) \pm |\DDelta(\rr)|$, mixing the spatially extended quasi-Bloch states with opposite pseudospins. Appendix \ref{sec_app_res_pot} shows that the adiabatic approximation indeed works less well with a larger $\psi$. A more formal treatment of deviation from the adiabatic approximation based on second-order perturbation theory is presented in Appendix \ref{sec_app_SW}.

\begin{figure}
	\centering
	\includegraphics[width=0.48\textwidth]{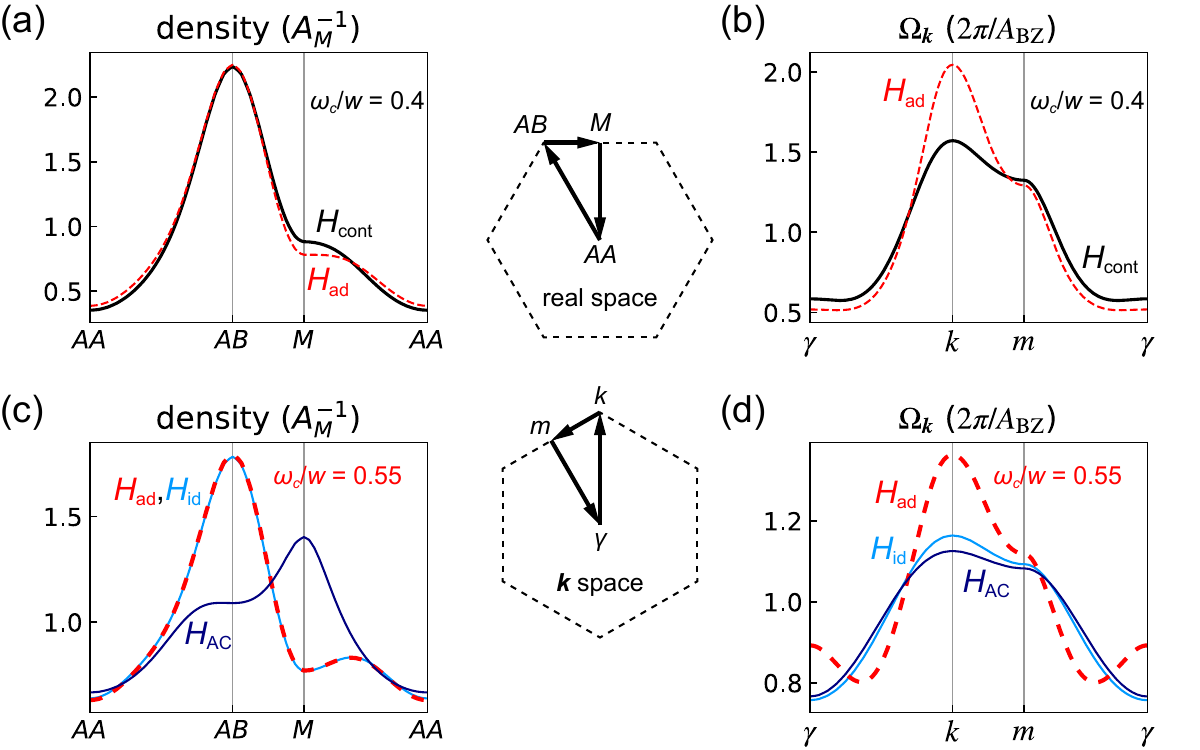}
	\caption{(a) The real-space charge density distribution of the filled first band of twisted bilayer \ce{WSe_2} with the twist angle $\omega_c/w = 0.4$, calculated using both the continuum model and the full adiabatic approximation. (b) The $\kk$-space Berry curvature distribution $\Omega_{\kk}$ of the first band of $\omega_c/w = 0.4$ \ce{WSe_2}, calculated under the same two models. (c)-(d) The same quantities as (a)-(b), now calculated for the adiabatic approximation band ($H_{\rm ad}$) under $\omega_c/w = 0.55$, the hypothetical ideal band fitted from the adiabatic charge density distribution ($H_{\rm id}$) at $\omega_c/w = 0.55$, and the AC band ($H_{\rm AC}$). All real-space and $\kk$-space plots are along the symmetric lines indicated in the middle panel. Each plotted curve is labeled with its corresponding model Hamiltonian with the same color. Note that in (c) the $H_{\rm ad}$ and $H_{\rm id}$ curves are identical within the widths of the plotted lines. The full 2D plots for these parameter values are presented in Fig. \ref{fig_2Dcomparisons_app} in the appendix.}
	\label{fig_comparisons2}
\end{figure}

Fig. \ref{fig_comparisons} (d) shows that the adiabatic approximation also reproduces the appearance of 
nearly ideal quantum geometry over a small range of twist angles, including and mostly above
those at which the bandwidth is minimized.
In the rest of the paper we will assume that adiabatic approximation results are converged with respect to 
Landau level and momentum space shell cutoffs at $N_{\rm LL} = 25$ and $N_{\rm shell} = 15$ in \ce{WSe_2} above $0.2^\circ$.
Figs. \ref{fig_comparisons2} (a)-(b) show that for $\omega_c = 0.4$ \ce{WSe_2}, the adiabatic approximation well reproduces both the real-space total charge density and the overall trend of momentum-space Berry curvature of the first valence moir\'e band, though with an overestimate on the Berry curvature peaks at mBZ corner points.

\section{Comparisons between Aharonov-Casher and Adiabatic Band Properties}
\label{sec_comp_AC_ad}

In Sec. \ref{subsec_AC}, we have pointed out that when the residual potential $U(\rr)$ in Eq. (\ref{eq_ad_Ham1}) vanishes
we are presented with a Bloch version of AC \cite{aharonov1979ground} states, and 
that the resulting AC bands have ideal quantum geometry. When the residual potential $U(\rr)$ is non-zero, the adiabatic
flat band generally develops dispersion and the Bloch state 
wave functions are altered by $\kk$-dependent mixing between the AC band and higher energy bands.
We expect that, because the flat band wavefunctions are smooth, 
band properties will be less sensitive to higher reciprocal lattice vector shells in the Fourier expansion of $U(\rr)$.
In particular, when the first (and most important) harmonic of the potential $U_1 = \Delta_{+,1} - \omega_c\xi_1$ (see Eq. (\ref{eq_ideal_pseudopotential})) is tuned to zero,
fractionally-filled-band many-body ground states are still expected to exhibit lowest-Landau-level like correlations.
This argument is supported by Fig. \ref{fig_theme1} in the introduction, from which we see that higher shells of $U(\rr)$ very weakly affect the bandwidth and the ideal quantum geometry when $U_1 = 0$.
(We do note that there seems to be a line, whose origin we do not understand at present,
in the $U_2-U_3$-plane along which the band remains particularly flat.)
We ascribe the relative insensitivity to higher Fourier components to the smoothness of wavefunctions in the 
lowest energy hole bands.
The twist angle at which $U_1$ vanishes corresponds to $\omega_{c0}^{U_1} = \Delta_{+,1}/\xi_1 \approx 0.52w$ under the parameter set of \ce{WSe_2} we focus on, which is close to the adiabatic approximation magic angle condition $\omega_{c0}^{\rm ad} \approx 0.55w$.
The approximate match between these two twist angles occurs over a wide region of the continuum model parameter space
as discussed in Appendix \ref{sec_app_AClimit}.

We have noticed from Fig. \ref{fig_comparisons} (d) that at the adiabatic approximation magic angle $\omega_{c0}^{\rm ad}/w \approx 0.55$ the 
quantum geometry of the adiabatic moir\'e band appears extremely close to ideal, whereas the continuum band is slightly less close. To illustrate the idealness of the adiabatic band, we construct a hypothetical ideal band that reproduces its full charge density, using an approximate method we will discuss shortly. The hypothetical ideal band indeed has very similar wave functions to the adiabatic band, as shown in Appendix \ref{sec_app_ad_id_wf}, but is very different from the AC band owing to strongly spatially varying $U(\rr)$.
In the rest of this section, we examine the influence of 
sub-leading Fourier components of the residual potential $U(\rr)$ on the first moir\'e band by comparing the charge density and Berry curvature distributions calculated from the AC band, the hypothetical ideal band and the full adiabatic approximation under this magic angle.

\subsection{Charge Density Distribution}
\label{sec_AC_ad_chargedensity}

In Fig. \ref{fig_comparisons2} (c) we plot the real-space charge density distribution of a full band in the three models.
The charge of the full AC band is concentrated at $M$ points (as labeled in Fig. \ref{fig_realspace} (a)).
This is a direct result from the distribution of effective magnetic field $\mcB(\rr)$: by the plasma analogy \cite{laughlin1983anomalous}, $\mcB(\rr)$ acts like a nonuniformly charged background with the charge opposite to the carrier, which gives an ``electrostatic potential'' $-\alpha(\rr)$ via $\nnabla^2\alpha = \mcB$.
Because of the sharp peaks in $|\mcB(\rr)|$ at the $M$ points, as shown in Fig. \ref{fig_realspace} (a), the periodic part of the electrostatic potential, $-\chi(\rr)$, has potential wells at those points, which attracts the band carrier charge via the factor $e^{\chi(\rr)}$ in the wave function form Eq. (\ref{eq_AC_wavefunction}).

Now, we examine the effect of $U(\rr)$. We see from Fig. \ref{fig_realspace} (b) and (c) that the two contributing parts of $U(\rr)$ in Eq. (\ref{eq_ideal_pseudopotential}), $\Delta_+(\rr)$ and $\xi(\rr)$, are peaked at different real-space points, which implies that $U(\rr)$ has high peaks, or deep potential wells in terms of holes, at the $AB$ and $BA$ points labeled in Fig. \ref{fig_realspace} (a).
This explains the high concentration of charge density at these points in the adiabatic approximation.
On this basis we expect that at a larger twist angle, the contribution of $\Delta_+(\rr)$, and hence the high peaks in $U(\rr)$, 
are smaller relative to the Landau-level separation $\omega_c$ and results in a more uniform charge distribution along the Wigner-Seitz cell edge.

At the adiabatic magic angle the extremely ideal quantum geometry of the adiabatic band suggests an ideal band approximation
\begin{equation}
    \psi_{\kk}^{\rm ad}(\rr) \approx \psi_{\kk}^{\rm id}(\rr) = e^{\chi^{\rm id}(\rr)} \psi_{\kk}^{\LLL}(\rr),
    \label{eq_psi_id}
\end{equation}
which can be viewed as an ``AC'' band with a different effective magnetic field $\mcB^{\rm id}(\rr) = \nnabla^2\chi^{\rm id}$. While there is no analytical way to solve for the exact form of $\chi^{\rm id}(\rr)$ given its total charge density distribution $\rho^{\rm id}(\rr) = \rho^{\rm ad}(\rr)$, we make a very accurate approximation that the charge density distribution has the same shape as $e^{2\chi^{\rm id}(\rr)}$, which is analytically demonstrated in Appendix \ref{sec_app_rho_Phi}. On this basis, we assume $e^{2\chi^{\rm id}(\rr)} \propto \rho^{\rm ad}(\rr)$, which is the adiabatic charge density distribution, and see from Fig. \ref{fig_comparisons2} (c) that the charge density of that ideal band is indeed nearly identical to that of the adiabatic band, and very different from that of the AC band.

\subsection{Berry Curvature Distribution}
\label{sec_AC_ad_Berry}

Fig. \ref{fig_comparisons2} (d) shows that both the AC band and the fitted ideal band have more
uniform Berry curvature distributions than the adiabatic band.
The overall smoothness and the shallow dip at the mBZ center can be easily understood from
Eq. (\ref{eq_Berrycurv_AC}) for a general AC (or ideal) band: because of the 
suppression \cite{morales-duran2024magic} provided by the 
$\lambda_{\GG}$ form factors, which take the values $0.163$, $4.33 \times 10^{-3}$ and $7.06 \times 10^{-4}$ on the first three shells of the reciprocal lattice, all higher-shell Fourier components of the Berry curvature variation are negligible.

In contrast, the full adiabatic approximation Berry curvature distribution has small but visible peaks at the $k$ and $\gamma$ points,
which is a signature of non-negligible second-shell Fourier component.
This indicates that it is impossible to find an ideal band that reproduces both the charge profile and the Berry curvature profile of the adiabatic band.
In fact, it seems unlikely that any exact ideal band could exactly
reproduce this Berry curvature distribution -- if such ideal band does exist, its $e^{2\chi(\rr)}$ will have an exponentially large second-shell Fourier component which easily violates the positive definiteness of $e^{2\chi(\rr)}$. We do attempt to apply the
inverse procedure of Eqs. (\ref{eq_Berrycurv_AC})
to the adiabatic Berry curvature, assuming it to be identical to that of some alternative AC band. We 
find that the violation of positive definiteness in $e^{2\chi(\rr)}$ persists upon including further shells of Fourier components until the exponentially growing numerical error turns everything meaningless. While it is not to exclude the possibility that very high spatial harmonics of its $e^{2\chi(\rr)}$ could form a strange shape that compensates for the negative parts introduced by the lower harmonics, in this case, $e^{2\chi(\rr)}$ would have exponentially high sharp peaks that are totally inconsistent with the charge density distribution of the adiabatic band.
Rather, we argue that the second-shell Fourier component of the Berry curvature has to mostly come from the deviation of the wave functions from the ideal band limit ($\delta\psi_{\kk}$). Simply speaking, the change in the idealness of quantum geometry, $1 - \Omega_{\kk}/\tr g_{\kk}$, is second order in $\delta\psi_{\kk}$, while the change in Berry curvature is first order. Details are provided in Appendix \ref{sec_app_ad_id_wf}.

\section{Landau-Level to Haldane-Model Crossover}
\label{subsec_otherbands}

\begin{figure}
	\centering
	\includegraphics[width=0.48\textwidth]{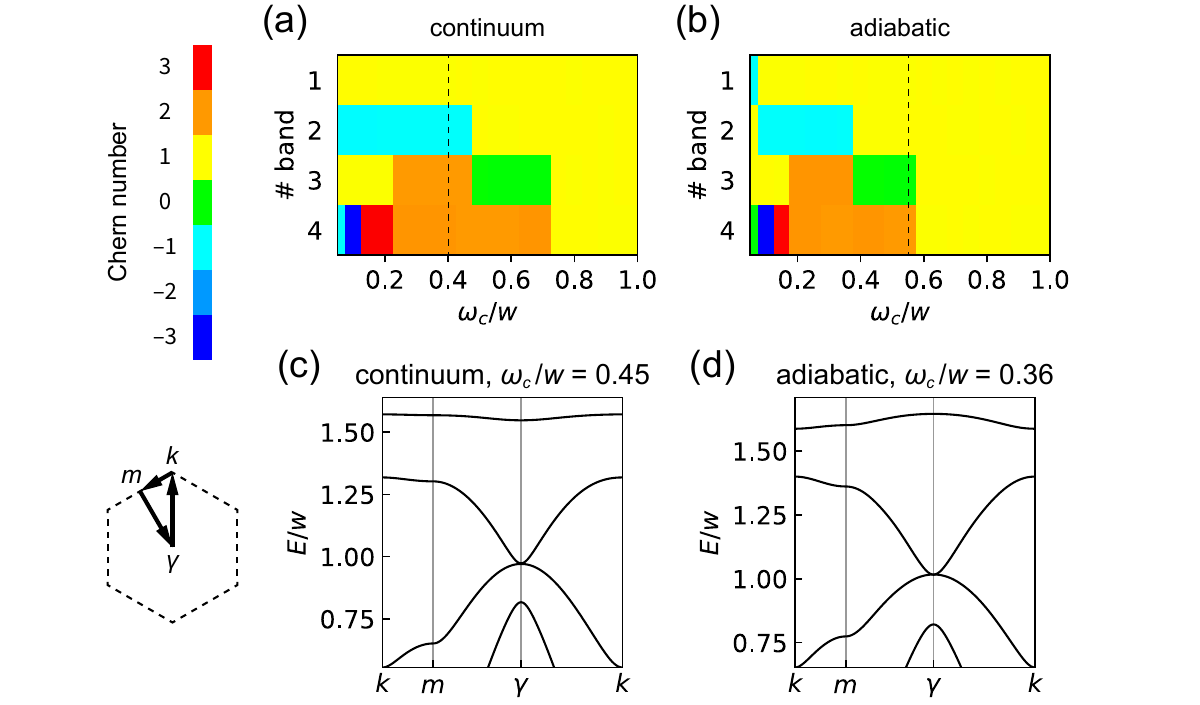}
	\caption{(a)-(b) Dependence of Chern numbers of the first 4 valence bands on the twist angle of twisted \ce{WSe_2} calculated under (a) the continuum model and (b) the adiabatic approximation. Different colors stand for different Chern numbers, as the color bar on the left shows. The precision in $\omega_c/w$ is 0.05 for both panels. The minimum-bandwidth magic twist angles are marked by vertical dashed lines. (c)-(d) The moir\'e band structures of the two models with parameters indicated on the top, along the path illustrated on the left.}
	\label{fig_Cherns}
\end{figure}

Figures \ref{fig_Cherns} (a) and (b) show that the twist angle dependence of the Chern numbers of the first few valence bands 
are well reproduced by the adiabatic approximation,
except at very small twist angles where the 25-Landau-level truncation is no longer sufficient for convergence.  The critical twist angles for topological phase transition are generally slightly underestimated. While the first valence band always has Chern number $\mcC_1 = 1$, the Chern number of the second band $\mcC_2$ transitions from $-1$ to $1$ at a critical twist angle that is larger than the magic angle in the continuum model 
but smaller than the magic angle in the adiabatic approximation. 
In both cases, the transition is triggered by a quadratic inversion between the second and the third bands at the $\gamma$ point with $4\pi$ Berry curvature transfer, as suggested by Figs. \ref{fig_Cherns} (c) and (d)

An effective Haldane model \cite{haldane1988model} (or Kane-Mele model \cite{kane2005topological, kane2005quantum}, considering both spins/valleys) describes \cite{wu2019topological, wang2023staggered, crepel2023anomalous, liu2024gatetunable} the first two bands 
in the small twist angle regime where they have opposite Chern numbers.  
In the continuum model, the Wannier orbitals corresponding to this two-band truncation 
sit on a honeycomb lattice with the $A$ and $B$ sublattices at the $AB$ and $BA$ stacking points in the moir\'e pattern.
The $A$ and $B$ sublattice Wannier orbitals are polarized toward the top and bottom layers.
Since the adiabatic approximation is accurate at small twist angles, we expect that it also captures the $\mcC_2 = -1$ regime
of the adiabatic approximation model.  In the adiabatic approximation, the layer polarization of localized Wannier orbitals
is encoded by the position-dependence of the effective magnetic field, which results in turn from the spatial variation of the 
local layer pseudospin fields which point in opposite directions at the $AB$ and $BA$ stacking points.

At larger twist angles, the Chern numbers of the first few bands are uniformly 1, which is a signature of a Landau-level-like structure, 
at least in a topological sense.  This crossover between a Haldane-model like regime at small twist angles and a 
Landau-level-like regime at large twist angles has also been predicted under the parameters of twisted homobilayer \ce{MoTe_2} \cite{qiu2023interactiondriven, xu2024maximally}.
Realization of a Landau-level-like regime paves the way to engineering even-denominator FCI states in a flat band resembling the $n=1$ Landau level of a 2D electron gas \cite{fujimoto2024higher, reddy2024nonabelian, xu2024multiple}. These states could support
non-Abelian anyons that are promising resources of topological quantum computation.

\section{Summary and Discussion}
\label{sec_conclusion}

The adiabatic approximation, in which the layer degree-of-freedom is removed and replaced by a non-uniform periodic effective magnetic field, is an attractive alternative to direct continuum model descriptions.
We have shown that with realistic model parameters \cite{devakul2021magic} near the magic angle, the adiabatic approximation indeed accurately reproduces the continuum model band properties \cite{wu2019topological}.    
It was first introduced \cite{morales-duran2024magic} to explain the appearance of 
ideal flat bands \cite{morales-duran2023pressureenhanced} in twisted homobilayer TMDs 
in terms of minimal mixing between a flat adiabatic-approximation lowest Landau level 
and higher Landau levels.
By introducing the concept of Aharonov-Casher bands, we have explicitly shown that nearly-ideal bands can 
emerge even with finite Landau level mixing.

In homobilayer TMD moir\'es, non-perturbative 
many-body methods are needed \cite{xu2024maximally, yu2024fractional, dong2023composite, abouelkomsan2024band, wang2024fractional} to study the competition between FCI and charge-density-wave states at fractional band fillings. The Landau level representation of the adiabatic approximation Hamiltonian provides 
an attractive basis for numerical many-body simulation methods like density matrix renormalization group (DMRG) and 
exact diagonalization (ED). By using Landau level basis states, it may be possible to achieve a deeper understanding of the essential physics of
this specific FCI system.

Research directed towards deriving the most accurate continuum model Hamiltonians for specific 
homobilayers is still in progress \cite{reddy2023fractional, xu2024maximally, jia2024moire, mao2024transfer, wang2024fractional, zhang2024polarizationdrivena}. In some cases the accuracy of adiabatic approximation can be limited 
by rapid spatial variations in the Hamiltonian or small magnitudes of the layer-pseudospin field.
The adiabatic approximation can be improved, we believe, by relaxing the strict 
locking invoked here between the layer pseudospin polarization function $u_s(\rr)$ and the layer pseudospin field,
allowing Bloch bands with spinors of the form
\begin{equation}
	\psi_{\kk,s}^{\rm cont}(\rr) = u_s(\rr) \psi_{\kk}^{\rm ad}(\rr),
\end{equation}
where $s$ is the layer index. Both $u_s(\rr)$ and $\psi_{\kk}^{\rm ad}(\rr)$ can in principle be obtained self-consistently 
using a variational approach or by directly fitting the continuum-model wave functions so that the effective magnetic field distribution in the adiabatic approximation is optimized. 
We expect that the optimized effective magnetic field will be smoother than in the original simple 
adiabatic approximation, giving a more accurate description of the moir\'e band properties that works
in a wider region of parameter space.

There is evidence from experiment \cite{cai2023signatures, zeng2023thermodynamic, park2023observation, xu2023observation}
that the Chern band character of TMD homobilayer moir\'e bands is most 
robust not near zero density, but near band filling $\nu=-1$. It may be that the ideal bands relevant to 
experiment are the electron and hole-like quasiparticle bands of the $\nu=-1$ states, which are 
influenced by Hartree and Fock self-energies.  Because the Fock self-energies are 
non-local, the simple adiabatic approximation can no longer be applied to this Hamiltonian, but 
the generalization imagined in the preceeding paragraph can still be applied.
The Hartree potential of the hole-filled bands at $\nu=-1$ will tend to smooth the layer 
skyrmion field, and may give rise to a more uniform effective magnetic field and a more uniform effective potential.
With smoother magnetic field, the adiabatic approximation is expected to converge at a 
smaller Landau-level cutoff, which would allow future DMRG and ED simulations to give more accurate results.

The adiabatic approximation can be directly generalized to multilayer systems. One example is symmetric twisted homotrilayer TMDs, 
which can be decoupled into an effective bilayer and an effective monolayer based on the mirror symmetry about the middle layer
in analogy to ABA twisted trilayer graphene \cite{li2019electronic, khalaf2019magic}.
Another example is the twisted homobilayer TMD proximitized with a third layer of a different type of TMD, 
in which the twist-angle-tuned interplay between hombilayer and heterobilayer moir\'e patterns,
analogous to a similar interplay in twisted bilayer graphene aligned with hexagonal boron nitride \cite{cea2020band, shi2021moire, shin2021electronhole},
could lead to interesting physics. Both types of systems could potentially host triangular and/or sublattice-staggered hexagonal lattices of atomic-like Wannier orbitals that could potentially lead to frustrated Hubbard models and to exotic strongly correlated phases upon doping, possibly including unconventional superconductivity. At the same time, flat ideal bands and FCI states are still in principle possible to engineer at certain band filling if the hexagonal sublattice-staggered potential is properly screened, pushing the physics picture towards the Haldane-model regime of twisted homobilayers. 
It may even be possible to engineer higher-Chern number ideal bands \cite{wang2022hierarchy, ledwith2022family, guerci2024chern, guerci2024nature, popov2023magic, foo2024extended} in multilayer TMDs, which have more interesting color-entangled wave functions \cite{dong2023manybody, wang2023origin} that can be mapped to multilayer Landau levels, and can be topologically classified based on the motion of zeros in their wave functions.

In this paper we have argued that the nearly ideal bands of homobilayers are related to the Aharonov-Casher states \cite{aharonov1979ground} of a 2D electron gas with $g$-factor $2$.  The cancellation between 
zero-point kinetic energy and Zeeman energy in the Aharonov-Casher case is replaced by a vanishing first-shell Fourier component of 
the residual potential $U(\rr)$.
We have shown that under this circumstance the effect of the a realistic 
residual potential $U(\rr)$ on the flat dispersion and ideal quantum geometry of the AC band is generally small,
despite the band mixing effects from the sub-leading Fourier shells. 
It will be interesting to seek a more complete analytical understanding of this property and to
explore under what conditions it still holds upon introducing the exchange potential and/or interaction-induced band mixing effects.

Based on our findings, we suggest that projection onto AC bands that includes only the first reciprocal lattice shell of the magnetic field can be used to construct a minimal model for the nearly ideal bands in TMD homobilayer moir\'e materials. In these models the charge is concentrated near, and relatively evenly distributed along, the Wigner-Seitz cell edge.
Corrections to the charge density distribution, band dispersion, and deviations from band ideality 
can be modeled by adding first-shell residual potentials and also
including second- and/or third-shell Fourier components of either 
the potential or the magnetic field.
We expect that phenomenological models in this class
will be able to describe all $C_6$-symmetric TMD moir\'e materials, 
and that many-body simulations on this model will be able to draw a relatively complete picture of the interplay between bandwidth, quantum geometry, charge-density distributions and other factors in 
the FCI states of these materials.

\begin{acknowledgments}
This research was supported by the US Department of Energy (DOE), Office of Science, Basic Energy Sciences (BES), under award numbers DE-SC0019481.
We gratefully acknowledge helpful discussions with Martin Claassen, Liang Fu, Matteo Ippoliti, Chong Wang, Yihang Zeng and Bo Zou. This work was enabled by computational resources provided by the Texas Advanced Computing Center. 
\end{acknowledgments}

\appendix
\renewcommand\thefigure{A\arabic{figure}}
\setcounter{figure}{0}

\begin{figure*}
	\centering
	\includegraphics[width=0.96\textwidth]{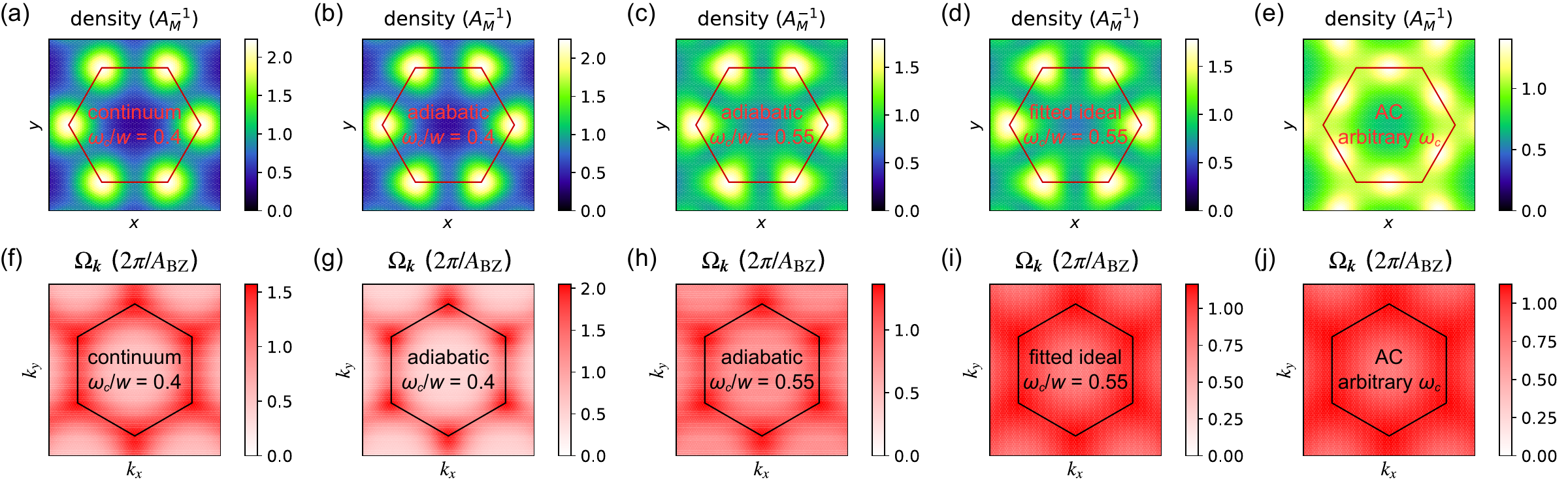}
	\caption{(a)-(e) 2D color plots of the real-space charge density of the filled first band calculated from the model parameters associated with \ce{WSe_2}: (a) the continuum model under the minimum-bandwidth magic twist angle $\omega_c/w = 0.4$, (b) the adiabatic approximation under the same twist angle, (c) the adiabatic approximation under the adiabatic magic angle $\omega_c/w=0.55$, (d) the ideal band fitted from the adiabatic charge density distribution under $\omega_c/w=0.55$ and (e) the AC band under arbitrary magic angle. (f)-(j) 2D color plots of the $\kk$-space Berry curvature under the same model and parameter settings as (a)-(e).}
	\label{fig_2Dcomparisons_app}
\end{figure*}

\begin{table*}
	\centering
	\setlength{\tabcolsep}{10pt}
	\begin{tabular}{|c|c|c|c|c|c|c|c|}
		\hline Reference & Material & $a$ (\AA) & $m$ ($m_e$) & $w$ (meV) & $V$ (meV) & $\psi$ ($^\circ$) & $V/w$ \\
		\hline \onlinecite{wu2019topological} & \ce{MoTe_2}, rigid & 3.472 & 0.62 & 8.5 & 8 & 89.6 & 0.94 \\
		\hline \onlinecite{reddy2023fractional} & \ce{MoTe_2}, relaxed & 3.52 & 0.62 & 13.3 & 11.2 & 91 & 0.84 \\
		\hline \onlinecite{xu2024maximally} & \ce{MoTe_2}, relaxed & 3.52 & 0.62 & 11.2 & 9.2 & 99 & 0.82 \\
		\hline \onlinecite{wang2024fractional} & \ce{MoTe_2}, relaxed and corrugated & 3.52 & 0.6 & 23.8 & 20.8 & 107.7 & 0.87 \\
		\hline \onlinecite{devakul2021magic} & \ce{WSe_2}, relaxed & 3.317 & 0.43 & 18 & 9 & 128 & 0.5 \\
		\hline \onlinecite{morales-duran2023pressureenhanced} & \ce{WSe_2}, relaxed & 3.297 & 0.337 & 8.9 & 6.4 & 115.7 & 0.72 \\ \hline
	\end{tabular}
	\caption{A compilation of model parameters obtained from first principle simulations in literature. $m_e$ is the free electron mass. Note that by making symmetry transforms to the system one can always make all the parameters positive, as discussed in Sec. \ref{subsec_cont}.}
	\label{table_app_modelparameters}
\end{table*}

\section{Adiabatic Approximation}
\label{sec_app_ad}

In this appendix section, we start from a generalized version of the continuum model for
%In the continuum model of a general
twisted homo-$N$-layer TMDs, where the valence bands are described by nonrelativestic holes moving under an $N \times N$ layer-pseudospin entangled potential. The holes have 2 spin-valley locked flavors and $N$ layer pseudospins. The projected Hamiltonian in valley $K$ (or spin up) writes
\begin{equation}
	H = -\frac{\hat\pp^2}{2m} \Id_N + \Delta(\rr),
	\label{eq_app_cont_Ham}
\end{equation}
where $m$ is the effective hole mass in the valence band of the material, $\Id_N$ is the $N \times N$ identity matrix and the position-dependent $N \times N$ matrix $\Delta(\rr)$ describes the moir\'e potential.

For the adiabatic approximation, we do a 
%can be generalized from two-level spins to $N$-component spinors. In both cases the goal is to simply account for real space Berry phases associated with spinor textures.
${\rm U}(N)$ gauge transformation on the continuum Hamiltonian Eq. (\ref{eq_app_cont_Ham}) that locally diagonalizes $\Delta(\rr)$:% yields
\begin{equation}
	\tilde H = \mcU^\dag H \mcU = -\frac{\left(\hat\pp\Id_N + \ffrakA(\rr)\right)^2}{2m} + \tilde\Delta(\rr),
    \label{eq_app_tildeHcont}
\end{equation}
where $\tilde\Delta(\rr) = \mcU^\dag(\rr) \Delta(\rr) \mcU(\rr)$ is a diagonal $N \times N$ matrix, $\mcU(\rr)$ is a space-dependent $N \times N$ unitary matrix, and
\begin{equation}
	\ffrakA(\rr) = -i\, \mcU^\dag(\rr) \nnabla \mcU(\rr)
\end{equation}
is the non-Abelian connection associated with the gauge transformation $\mcU$.
The $s$th diagonal element of $\tilde H$ is
\begin{equation}
	\tilde H_{ss} = -\frac{\left( \hat\pp + \ffrakA_{ss}(\rr) \right)^2}{2m} - \frac{\mcD_s(\rr)}{2m} + \Delta_s(\rr),
	\label{eq_app_tildeHcont_ss}
\end{equation}
where $\Delta_s(\rr)$ is the $s$th eigenvalue of $\Delta(\rr)$, and the extra term 
\begin{equation}
	\mcD_s(\rr) = \sum_{s'\ne s} \ffrakA_{ss'}(\rr) \cdot \ffrakA_{s's}(\rr)
\end{equation}
comes from the off-diagonal elements of $\ffrakA(\rr)$.
Denote the $s$th column vector of $\mcU(\rr)$ as $\uu_s(\rr)$, so that
\begin{equation}
	\ffrakA_{ss'}(\rr) = -i\uu_s^\dag \nnabla\uu_{s'} = i \left( \nnabla\uu_s^\dag \right) \uu_{s'},
\end{equation}
hence
\begin{equation} \begin{array}{c}
	\mcD_s(\rr) = \sum_{s'\ne s} \left( \nnabla\uu_s^\dag \right) \uu_{s'} \cdot \uu_{s'}^\dag \nnabla\uu_s \\\addlinespace
	= \nnabla\uu_s^\dag \cdot \nnabla\uu_s - \left( \nnabla\uu_s^\dag \right) \uu_s \cdot \uu_s^\dag \nnabla\uu_s,
\end{array} \end{equation}
which is exactly the trace of quantum metric of the $s$th eigenstate manifold of $\Delta(\rr)$.

If we are interested only on the highest energy valence band 
states we can truncate to the highest-eigenstate manifold.
For the $N=2$ case appropriate to homobilayers,
we label this as ``$+$'' so that $H_{\rm ad} = \tilde H_{++}$. 
In this case, Eq. (\ref{eq_app_tildeHcont_ss}) becomes Eq. (\ref{eq_ad_Ham0}) in the main text, where $\mmcA(\rr) = \ffrakA_{++}(\rr)$ and $Z(\rr)$ is given by Eq. (\ref{eq_Z(r)}). Now we have $\Delta_+(\rr) = \Delta_0(\rr) + \Abs{\DDelta(\rr)}$ and it can be shown, regardless of the $\rm U(1)$ gauge freedom remaining after 
rotating the pseusospins to alignment, that the effective magnetic field $\mcB(\rr)$ and the function $\mcD(\rr) = \mcD_+(\rr)$ are given respectively by
\begin{equation}
	\mcB(\rr) = \nnabla \times \mmcA(\rr) = \frac{\nn \cdot (\partial_x\nn \times \partial_y\nn)}{2}
\end{equation}
and
\begin{equation}
	\mcD(\rr) = \ffrakA_{+-}(\rr) \cdot \ffrakA_{-+}(\rr) = \frac{|\partial_x\nn|^2 + |\partial_y\nn|^2}{4},
\end{equation}
where $\nn(\rr) = \DDelta(\rr) / |\DDelta(\rr)|$ and ``$-$'' is the label of the lower eigenstate of $\Delta(\rr)$.

\section{Relationship Between Model Parameters and Residual Potential}
\label{sec_app_res_pot}

\begin{figure*}
	\centering
	\includegraphics[width=0.96\textwidth]{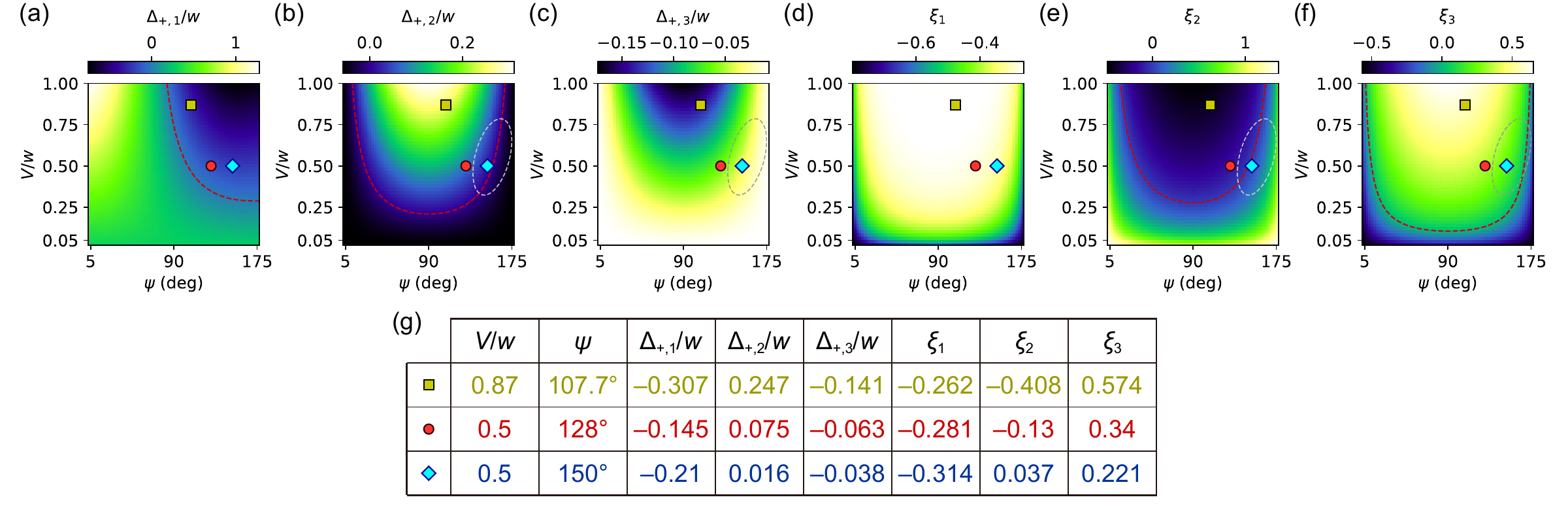}
	\caption{(a)-(f) Maps of the first three harmonic shells of Fourier components of (a)-(c) $\Delta_+(\rr)/w$ and (d)-(f) $\xi(\rr)$, the two contributions to the residual potential $U(\rr)$ in Eq. (\ref{eq_ideal_pseudopotential}). 
The red dashed lines are the zero contour lines. The positions of the \ce{MoTe_2} parameter set $(V/w, \psi) = (0.87, 107.7^\circ)$,
the \ce{WSe_2} parameter set $(0.5, 128^\circ)$ and the extra parameter set $(0.5, 150^\circ)$
are respectively marked by square, circle and rhombus in each channel.
The region where $\omega_{c0}^{U_1} = \Delta_{+,1}/\xi_1 > 0$ and $|\Delta_{+,i}|$ and $|\xi_i|$ are small for both $i=2$ and $3$ are roughly indicated by dashed grey circles in (b), (c), (e) and (f).
(g) The table of local values of the plotted quantities in (a)-(f) at the marked points.
}
	\label{fig_Deltaxi}
\end{figure*}

\begin{figure}
	\centering
	\includegraphics[width=0.48\textwidth]{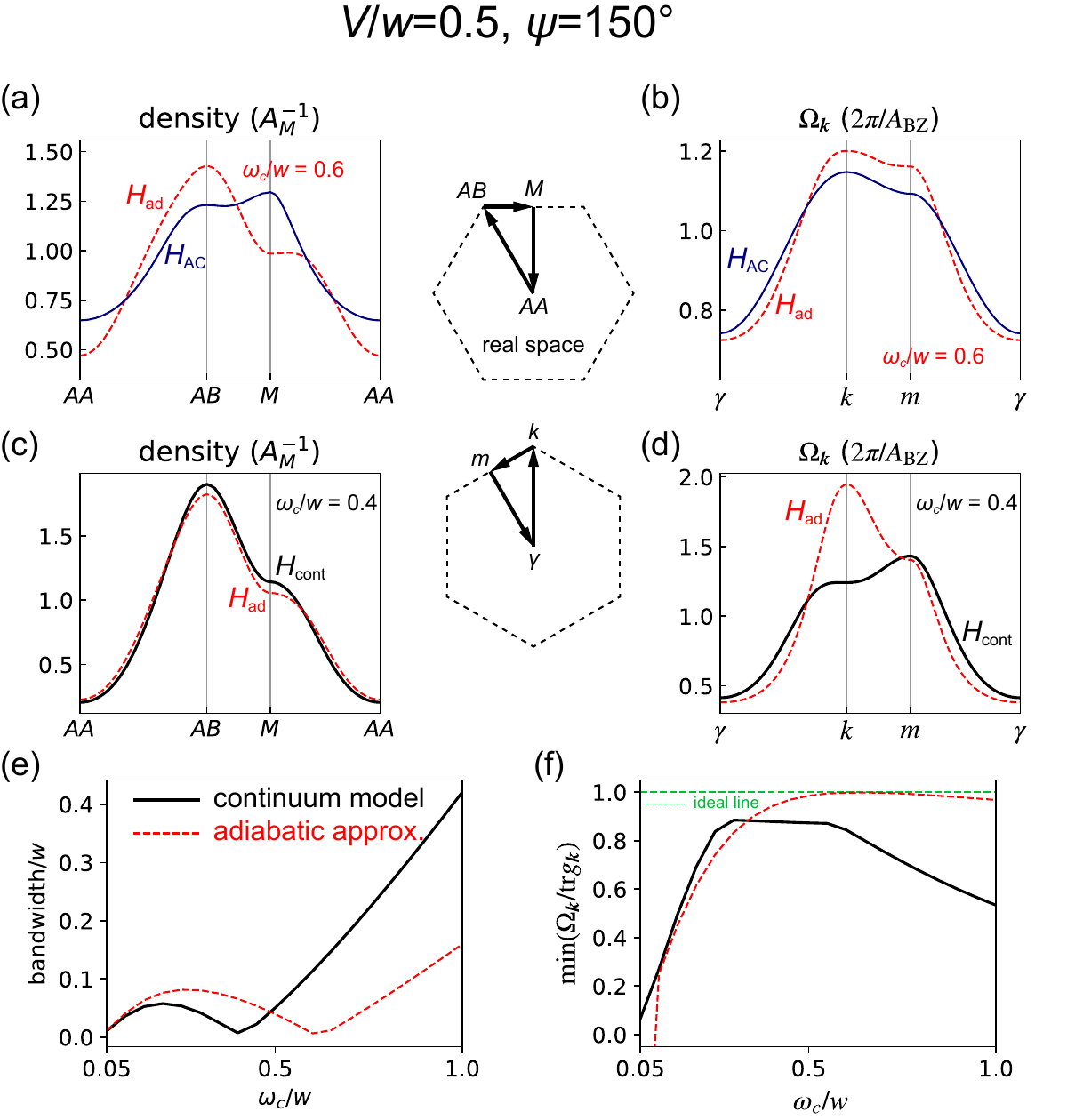}
	\caption{(a) The real-space charge density distribution of the filled first band, and (b) the $\kk$-space Berry curvature distribution $\Omega_{\kk}$ of the first band of the adiabatic approximation and the AC band derived from the system under continuum model parameters $(V/w, \psi) = (0.5, 150^\circ)$, under the twist angle $\omega_c/w = 0.6$, which is near the adiabatic magic angle.
	(c)-(d) The same quantities as (a)-(b), now calculated for the continuum model and adiabatic approximation bands under $\omega_c/w = 0.4$, which is near the continuum magic angle. All real-space and $\kk$-space plots are along the symmetric lines indicated in the middle panel.
	(e) The bandwidth and (f) the quantum geometry idealness characterized by $\min(\Omega_{\kk}/\tr g_{\kk})$ of the first band of the system under $(V/w, \psi) = (0.5, 150^\circ)$, as a function of twist angle, calculated under both the continuum model and the adiabatic approximation.
	In all adiabatic approximation results, the convergent Landau level and reciprocal space shell truncations $(N_{\rm LL}, N_{\rm shell}) = (25, 15)$ are used, as in the main text.}
	\label{fig_comparisons_app}
\end{figure}

The first few Fourier components of $\Delta_{+}(\rr)$ and $\xi(\rr)$, the two parts of the residual potential $U(\rr)$ 
as defined in Eq. (\ref{eq_ideal_pseudopotential}),
are presented in Fig. \ref{fig_Deltaxi} as maps of the continuum model parameter space.
The points in the parameter space that correspond to the explicit parameter sets on which we focus are 
marked on these plots and give a sense of the part of the space that is most experimentally relevant.
We see that
since $\xi_1$ is negative definite because the effective magnetic field is peaked on the
Wigner-Seitz cell boundary, the favorable region for Landau level like physics is the part of the parameter space in which $\Delta_{+,1}$ is negative, so that $U_1$ can be tuned to zero at some particular twist angle $\omega_{c0}^{U_1} = \Delta_{+,1}/\xi_1$.
In this region both the signs of $\Delta_{+,i}$ and $\xi_{i}$ are generally opposite for both the $i=2$ and 
the $i=3$ sub-leading shells, implying that they cannot cancel.  
Nevertheless, the deviation from the AC limit is expected to be smaller within the narrow region of the phase diagram  
near where $\Delta_{+,i}$ and $\xi_i$ are small for both $i=2$ and $i=3$ indicated in Fig.~\ref{fig_Deltaxi}.

To verify this argument, we take an example parameter set $(V/w, \psi) = (0.5, 150^\circ)$ from this region and see from Fig. \ref{fig_comparisons_app} (a)-(b) that for this parameter set, the AC band does reproduce the wave function of the first adiabatic band
well -- at least better than in the \ce{WSe_2} case discussed in the main text (see Figs. \ref{fig_comparisons2} (c)-(d)).
However, the rest of Fig. \ref{fig_comparisons_app} shows that for this parameter set, the adiabatic approximation reproduces the continuum model band properties less well than in the \ce{WSe_2} case (only the charge density distribution is well reproduced; See Figs. \ref{fig_comparisons} (a), (d) and \ref{fig_comparisons2} (a)-(b) for comparison with \ce{WSe_2}). Also, the quantum geometry of the first continuum-model band here is generally less ideal than in \ce{WSe_2}, suggesting that systems near this parameter regime is \textbf{not} a better candidate than materials like \ce{WSe_2} to realize FCI.

\section{Derivation of Landau-Level-Basis Hamiltonian Elements}
\label{sec_app_LL}

We start our derivation by substituting Eq. (\ref{eq_psikLLL}) and the expression of Landau-level raising operator, $a_0^\dag = (\ell/\sqrt{2}) (p^\dag + \mcA_0^*(\rr)) = (i/2\sqrt{2}\ell) (z^* - 4\ell^2\partial_z)$, to Eq. (\ref{eq_psik_nLL_raisedbya0}) in the main text:
\begin{widetext}
	\begin{equation} \begin{aligned}
		\psi_{\kk}^{n\rm LL}(\rr) &= \frac{1}{\sqrt{n!} (2\sqrt{2}\ell)^n} (iz^* - 4i\ell^2\partial_z)^n \left(
		e^{\frac{i}{2} \kk\cdot\rr} \psi_{\0}^{\LLL} (\rr + \ell^2\nn_z\times\kk) \right) \\
		&= \frac{e^{\frac{i}{2} \kk\cdot\rr}}{\sqrt{n!} (2\sqrt{2}\ell)^n} (iz^* + \ell^2k^* - 4i\ell^2\partial_z)^n \,
		\psi_{\0}^{\LLL} (\rr + \ell^2\nn_z\times\kk) \\
		&= \frac{e^{\frac{i}{2} \kk\cdot\rr}}{\sqrt{n!} (2\sqrt{2}\ell)^n} (iz^* + \ell^2k^* - 4\partial_k)^n \,
		\psi_{\0}^{\LLL} (\rr + \ell^2\nn_z\times\kk) \\
		&= \frac{e^{i\kk\cdot\rr}}{\sqrt{n!} (2\sqrt{2}\ell)^n} (\ell^2k^* - 4\partial_k)^n u_{\kk}^{\LLL}(\rr),
	\end{aligned} \label{eq_app_psik_nLL} \end{equation}
\end{widetext}
where
\begin{equation}
	u_{\kk}^{\LLL}(\rr) = e^{-i\kk\cdot\rr} \psi_{\kk}^{\LLL}(\rr) = e^{-\frac{i}{2} \kk\cdot\rr} \psi_{\0}^{\LLL} (\rr + \ell^2\nn_z\times\kk).
\end{equation}
Next, we define the notation
\begin{equation}
	\mcI_{\kk\pp}^{\GG} = \Braoperket{u_{\kk}^{\LLL}}{e^{i\GG\cdot\rr}}{u_{\pp}^{\LLL}},
	\label{eq_app_mcI}
\end{equation}
which can be calculated with the assist of the formula for the LLL wave function form factor \cite{wang2021exact}
\begin{equation}
	\Braket{u_{\kk}^{\LLL}}{u_{\kk+\qq}^{\LLL}} = e^{\ell^2\left( \frac{i}{2}\kk\times\qq - \frac{\qq^2}{4} \right)}
\end{equation}
and the $\kk$-space quasiperiodicity
\begin{equation}
	u_{\kk+\GG}^{\LLL}(\rr) = \overline\eta_{\GG} e^{-i\GG\cdot\rr + \frac{i\ell^2}{2} \GG\times\kk} u_{\kk}^{\LLL}(\rr),
\end{equation}
where $\overline\eta_{\GG}$ is the parity of the reciprocal lattice vector $\GG$ as described just after Eqs. (\ref{eq_Berrycurv_AC}). The ultimate expression of $\mcI_{\kk\pp}^{\GG}$ is
\begin{equation}
	\mcI_{\kk\pp}^{\GG} = \overline\eta_{\GG} e^{\frac{\ell^2}{2} \left( (k+G)^*p - k^*G \right) - \frac{\ell^2}{4} (\GG^2 + \kk^2 + \pp^2)}.
    \label{eq_app_mcIresult}
\end{equation}

In calculating the matrix element of $U'(\rr)$ in Eq. (\ref{eq_ad_Ham2}) between Landau levels, we use the following equation:
\begin{widetext}
	\begin{equation}
		\begin{aligned}
\left( \ell^2k - 4\partial_{k^*} \right)^n \left( \ell^2p^* - 4\partial_p \right)^{n'} \mcI_{\kk\pp}^{\GG}  
&= (-1)^n \left( 2\sqrt{2}\ell \right)^{n+n'} \overline\eta_{\GG} e^{\frac{\ell^2}{2} \left( (k+G)^*p - k^*G \right) - \frac{\ell^2}{4} \left( \GG^2 + \kk^2 + \pp^2 \right)} \\\addlinespace
&\times \mcL_{nn'} \biggl( \frac{\ell}{\sqrt{2}} (G+k-p)^*, \, \frac{\ell}{\sqrt{2}} (G+k-p) \biggr)
\end{aligned}
		\label{eq_app_mcI_deriv_formula}
	\end{equation}
	to obtain
	\begin{equation} \begin{array}{c}
		\Braoperket{\psi_{\kk}^{n\rm LL}}{U'(\rr)}{\psi_{\kk}^{n'\rm LL}} = \frac{1}{\sqrt{n!n'!} (2\sqrt{2}\ell)^{n+n'}} \sum_{\GG} U_{\GG}'
		\left. (\ell^2k - 4\partial_{k^*})^n \, (\ell^2p^* - 4\partial_p)^{n'} \mcI_{\kk\pp}^{\GG} \right|_{\pp=\kk} \\\addlinespace
		= \frac{(-1)^n}{\sqrt{n!n'!}} \sum_{\GG} U_{\GG}' \overline\eta_{\GG} \mcL_{nn'} \biggl( \frac{\ell G^*}{\sqrt{2}}, \frac{\ell G}{\sqrt{2}} \biggr)
		e^{\ell^2 \left( i\GG\times\kk - \frac{\GG^2}{4} \right)},
	\end{array} \label{eq_app_U'_betweenLLs} \end{equation}
\end{widetext}
where $\mcL_{nn'}(x, y) = e^{xy} \partial_x^n \partial_y^{n'} e^{-xy}$. The contribution of the other parts of $H'$ can be obtained by noting that from the relation between $\hat\Pi_0$ and $a_0$,
\begin{equation}
	\Braoperket{\psi_{\kk}^{n\rm LL}}{\hat\Pi_0^\dag \mcA'(\rr)}{\psi_{\kk}^{n'\rm LL}} = \frac{\sqrt{2n}}{\ell}
	\Braoperket{\psi_{\kk}^{(n-1)\rm LL}}{\mcA'(\rr)}{\psi_{\kk}^{n'\rm LL}},
\end{equation}
and by rewriting Eq. (\ref{eq_app_U'_betweenLLs}) in terms of $\mcA'(\rr)$. All combined gives Eq. (\ref{eq_LLformulation_Hamelement}) in the main text.

\section{Berry Curvature and Quantum Geometry Under Landau level Basis}
\label{sec_app_Berry_LLbasis}

We use \cite{wu2020quantum, abouelkomsan2023quantum}
\begin{equation}
	\ln \Braket{u_{\kk-\frac{\qq}{2}}}{u_{\kk+\frac{\qq}{2}}} = i\qq\cdot\mmcR(\kk) - \frac{\qq \cdot g\qq}{2} + \mcO(\qq^3)
\end{equation}
to approximate the quantum geometry on a $48 \times 48$ mesh in the mBZ, where $\mmcR(\kk) = -i \Braket{u_{\kk}}{\nnabla_{\kk} u_{\kk}}$ is the Berry connection and $g$ is the $2 \times 2$ Fubini-Study metric. In particular, the average Berry curvature in the triangle $\{ \kk, \kk\pm\qq_1, \kk\pm\qq_2 \}$ is obtained by accumulating the Berry phase along its perimeter, and the average Fubini-Study metric is solved from the obtained values of $\qq\cdot g\qq/2$ along the three edges, where $\qq_j = \GG_j/48$. This approach is used to calculate both real-space and momentum-space quantum geometries, and in theory always gives exact integer values of Chern number \cite{fukui2005chern}.

For a band with wave functions expressed upon the Landau level basis
\begin{equation}
	\Ket{u_{\kk}} = \sum_{n=0}^{N_{\rm LL}-1} c_{n\kk} \Ket{u_{\kk}^{n\rm LL}}
\end{equation}
where according to Eq. (\ref{eq_app_psik_nLL})
\begin{equation}
	u_{\kk}^{n\rm LL}(\rr) = e^{-i\kk\cdot\rr} \psi_{\kk}^{n\rm LL}(\rr) =
	\frac{(\ell^2 k^* - 4\partial_k)^n u_{\kk}^{\LLL}(\rr)}{\sqrt{n!} (2\sqrt{2}\ell)^n},
\end{equation}
the form factor $\Braket{u_{\kk}}{u_{\kk+\qq}}$ can be obtained from the Landau level form factor
\begin{equation}
	\Braket{u_{\kk}^{n\rm LL}}{u_{\pp}^{n'\rm LL}} = \frac{(\ell^2k - 4\partial_{k^*})^n (\ell^2p^* - 4\partial_p)^{n'} \mcI_{\kk\pp}^{\0}}
	{\sqrt{n!n'!} (2\sqrt{2}\ell)^{n+n'}},
\end{equation}
where the notation $\mcI$ is defined in Eq. (\ref{eq_app_mcI}). By formula Eq. (\ref{eq_app_mcI_deriv_formula}), we get
\begin{equation}
	\Braket{u_{\kk}^{n\rm LL}}{u_{\kk+\qq}^{n'\rm LL}} = \frac{(-1)^{n'}}{\sqrt{n!n'!}} \mcL_{nn'}
	\left( \frac{\ell q^*}{\sqrt{2}}, \frac{\ell q}{\sqrt{2}} \right) e^{\ell^2\left( \frac{i}{2}\kk\times\qq - \frac{\qq^2}{4} \right)}.
\end{equation}

\section{Beyond Adiabatic Approximation: Perturbation Theory Treatment}
\label{sec_app_SW}

In this section we attempt a perturbative estimate of the bounds of the twist-angle regime where the 
adiabatic approximation is accurate.  
We employ second-order perturbation theory to address the difference between the adiabatic approximation and the continuum model. 
For $N=2$ the gauge-transformed continuum model Hamiltonian (Eq. (\ref{eq_app_tildeHcont})) is given explicitly by 
\begin{widetext}
	\begin{equation}
		\tilde H = -\frac{1}{2m} \begin{pmatrix}
			\left( \hat\pp + \ffrakA_{++}(\rr) \right)^2 + \mcD(\rr)  &  \hat\pp\cdot\ffrakA_{+-} + \ffrakA_{+-}\cdot\hat\pp + \ffrakA_{+-}\cdot\tr\ffrakA \\
			\hat\pp\cdot\ffrakA_{-+} + \ffrakA_{-+}\cdot\hat\pp + \ffrakA_{-+}\cdot\tr\ffrakA  &  \left( \hat\pp + \ffrakA_{--}(\rr) \right)^2 + \mcD(\rr)
		\end{pmatrix} + \begin{pmatrix} \Delta_+(\rr) & 0 \\ 0 & \Delta_-(\rr) \end{pmatrix},
	\end{equation}
\end{widetext}
where we have used $\pm$ to label the rotated pseudospin up/down components,
$\mcD(\rr) = \mcD_+(\rr) = \mcD_-(\rr)$ and $\Delta_\pm(\rr) = \Delta_0(\rr) \pm \Abs{\DDelta(\rr)}$.
Locking the ${\rm U}(1)$ gauges of the two pseudospin sectors by the pseudo-time-reversal symmetry $\uu_-(\rr) = i\sigma^y \uu_+^{*}(\rr)$ guarantees that $\tr\ffrakA = 0$ and that $\ffrakA_{+-}(\rr)$ is bound in space. For any eigenstate $\ket{\tilde\psi_+}$ of the adiabatic approximation Hamiltonian $H_{\rm ad} = \tilde H_{++}$ with eigenvalue $\tilde E_+$, the energy shift due to virtual occupation in the pseudospin-down subspace can be captured by a Schrieffer-Wolf-like transformation:
\begin{equation}
	\delta E = \Braoperket{\tilde\psi_+}{\tilde H_{+-} \left( \tilde E_+ - \tilde H_{--} \right)^{-1} \tilde H_{-+}}{\tilde\psi_+}.
\end{equation}
From a semiclassical point of view, both $\hat\pp$ and $\ffrakA_{-+}(\rr)$ scale linearly with the twist angle $\theta$, while $(\tilde E_+ - \tilde H_{--})^{-1} \sim (\Delta_+ - \Delta_-)^{-1}$ does not scale with $\theta$, which tells us that $\delta E$ ($\delta E/\omega_c$) scales with $\theta^4$ ($\theta^2$), indicating breakdown of the adiabatic approximation by large $\delta E/\omega_c$ when the twist angle exceeds a certain limit.

To estimate $\delta E$, we employ a simplified formulation where the pseudospin-up state is taken as an LLL state:
\begin{widetext}
	\begin{equation}
		\delta E_{\kk} = \frac{\Braoperket{\psi_{\kk}^{\LLL}}{\left( \hat\pp\cdot\ffrakA_{+-} + \ffrakA_{+-}\cdot\hat\pp \right)
		\left( E_+ - \tilde H_{--} \right)^{-1}
		\left( \hat\pp\cdot\ffrakA_{-+} + \ffrakA_{-+}\cdot\hat\pp \right)}{\psi_{\kk}^{\LLL}}}{4m^2}.
		\label{eq_app_SW_k}
	\end{equation}
\end{widetext}
Next we replace $\hat\pp\cdot\ffrakA_{-+} + \ffrakA_{-+}\cdot\hat\pp$ with $\hat{\overline\PPi}_0\cdot\ffrakA_{-+} + \ffrakA_{-+}\cdot\hat\PPi_0$, where $\hat{\overline\PPi}_0 = \hat\pp - \mmcA_0$ is the time-reversal counterpart of $\hat\PPi_0 = \hat\pp + \mmcA_0$, $\mmcA_0$ is the vector potential of the uniform part of the effective magnetic field in the pseudospin-up subspace.
It can be shown from the quasiperiodicity of $\ffrakA_{-+}(\rr)$ that $\ffrakA_{-+}(\rr)$ acting on any magnetic translational eigenstate in the pseudospin-up subspace yields a magnetic translational eigenstate with the same quasimomentum $\kk$ in the pseudospin-down subspace, whose effective magnetic field is exactly the opposite of the pseudospin-up one. In addition, $\hat\PPi_0$ is associated with the Landau level raising and lowering operators by $\hat\Pi_{0x} + i\hat\Pi_{0y} = \hat\Pi_0 = \sqrt{2}\hat a_0/\ell$ (see Sec. \ref{subsec_LL}), which does not change the quasiperiodicity of the pseudospin-up magnetic translational eigenstate. Similarly, $\hat{\overline\PPi}_0$ is associated with the Landau level raising and lowering operators in the pseudospin-down subspace. Hereby we have proved that $\hat{\overline\PPi}_0\cdot\ffrakA_{-+} + \ffrakA_{-+}\cdot\hat\PPi_0$ conserves the quasimomentum while switching the pseudospin, which justifies replacing the middle Green's function in Eq. (\ref{eq_app_SW_k}) with its quasimomentum-$\kk$ sector.

Now we take a further simplification that replaces the Green's function with that of a uniform Landau level system with a global energy downshift from the pseudospin-up one by the indirect spatial gap $\Delta_g = \min \Delta_+(\rr) - \max \Delta_-(\rr)$:
\begin{equation}
	\Braoperket{\tilde\psi_{-,\kk}^{n\rm LL}}{\left( E_+ - \tilde H_{--} \right)^{-1}}{\tilde\psi_{-,\kk}^{n'\rm LL}} \sim
	\frac{\delta_{nn'}}{\Delta_g + n\omega_c},
\end{equation}
where $\tilde\psi_{-,\kk}^{n\rm LL}(\rr) = \tilde\psi_{+,(-\kk)}^{n\rm LL*}(\rr) = \psi_{-\kk}^{n\rm LL*}(\rr)$ is the wave function of the $n$th Landau level in the pseudospin-down subspace with quasimomentum $\kk$, which is the complex conjugate of that of the $n$th Landau level in the pseudopin-up subspace with quasimomentum $-\kk$. We truncate the Green's function within the first $N_{\rm LL}' = 50$ Landau levels and get the formula
\begin{equation}
	\delta E_{\kk} = \sum_{n=0}^{N_{\rm LL}'-1} \frac{\Abs{\Braoperket{\tilde\psi_{-,\kk}^{n\rm LL}}
	{\left( \hat{\overline\PPi}_0 \cdot \ffrakA_{-+} + \ffrakA_{-+} \cdot \hat\PPi_0 \right)}
	{\psi_{\kk}^{\LLL}}}^2} {4m^2 \left( \Delta_g + n\omega_c \right)},
\end{equation}
in which the numerator is numerically computed for each $(n, \kk)$ and some specific twist angle by summing up $48 \times 48$ real-space sample points, and obtained for other twist angles by the $\theta^4$ (or $\omega_c^2$) scaling law. The ``overall'' energy shift $\delta E$ is taken as the maximum of $\delta E_{\kk}$ over the mBZ.

\begin{figure}
    \centering
    \includegraphics[width=0.5\textwidth]{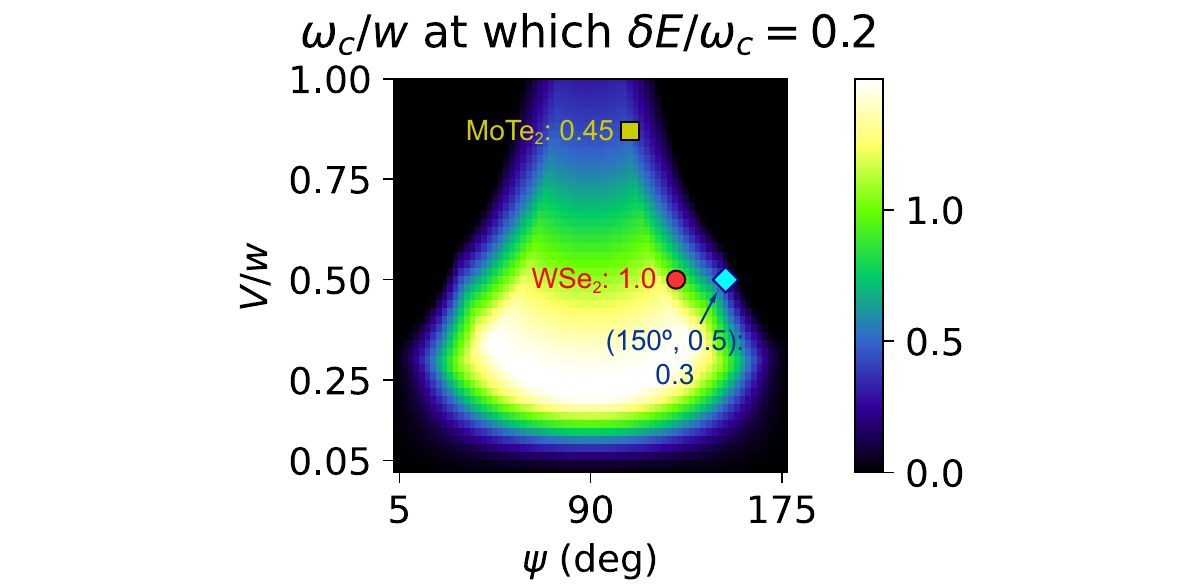}
    \caption{The estimated $\omega_c$ of the threshold twist angle under which the adiabatic approximation is expected to work well according to our second-order perturbation theory treatment, as a map of the continuum model parameters. The positions of \ce{MoTe_2} and \ce{WSe_2} parameter sets and the extra parameter set $(0.5, 150^\circ)$ are respectively marked by square, circle and rhombus, with their local values of estimated threshold $\omega_c/w$ marked on the side.}
    \label{fig_SW}
\end{figure}

Though the perturbation theory becomes inaccurate for small or negative $\Delta_g$, we do expect in other cases that $\delta E_{\kk}$ qualitatively well captures the dependence of deviation from the adiabatic approximation on the model parameter and twist angle. When $\Delta_g > 0$, we define the ``threshold'' twist angle below which the adiabatic approximation is good as the twist angle at which $\delta E/\omega_c = 0.2$. Otherwise, we specify that this threshold twist angle is 0. The map of threshold $\omega_c/w$ over the model parameters is shown in Fig. \ref{fig_SW}. Our formulation predicts a parameter window of $V/w$ between $0.2\sim 0.5$ and $\psi$ between $30^\circ\sim 150^\circ$, within which a relatively wide range of twist angle is accepted for good adiabatic approximation. The map correctly captures our previous observation that the adiabatic approximation gives a better estimate of the magic angle with the \ce{WSe_2} parameter set than with the \ce{MoTe_2} parameter set or the extra parameter set $(V/w, \psi) = (0.5, 150^\circ)$. It also reproduces the tendency of adiabatic approximation to break down at side values of $V/w$ and $\psi$ as we have argued in Sec. \ref{sec_comp_ad_cont}.

\section{AC Limit of the Adiabatic Approximation}
\label{sec_app_AClimit}

Here we approximate the dispersion of the first band by projecting the potential $U(\rr)$ into the AC ideal band subspace. From Eq. (\ref{eq_AC_wavefunction}) we have
\begin{equation} \begin{array}{c}
	E_{\kk}^{\rm AC} = \frac{\Braoperket{\psi_{\kk}^{\rm AC}}{U(\rr)}{\psi_{\kk}^{\rm AC}}}{\Braket{\psi_{\kk}^{\rm AC}}{\psi_{\kk}^{\rm AC}}}
	= \frac{\Braoperket{\psi_{\kk}^{\LLL}}{e^{2\chi(\rr)}U(\rr)}{\psi_{\kk}^{\LLL}}}{\Braoperket{\psi_{\kk}^{\LLL}}{e^{2\chi(\rr)}}{\psi_{\kk}^{\LLL}}} \\
	\addlinespace
	= \frac{\sum_{\GG} \overline\eta_{\GG} \lambda_{\GG} \tilde U_{\GG} e^{i\ell^2\GG\times\kk}}
	{\sum_{\GG} \overline\eta_{\GG} \lambda_{\GG} \Phi_{\GG} e^{i\ell^2\GG\times\kk}},
\end{array} \end{equation}
where $\tilde U_{\GG}$ and $\Phi_{\GG}$ are respectively the Fourier components of $e^{2\chi(\rr)} U(\rr)$ and $e^{2\chi(\rr)}$. 
(See Eq. (\ref{eq_LLformulation_Hamelement}) and associated definitions.)  
Assuming that the exponentially decaying nature of $\lambda_{\GG}$ allows us to truncate the reciprocal space summation after the first shell:
\begin{equation}
	E_{\kk}^{\rm AC} \approx E_{\kk}^{\rm AC1} = \frac{\tilde U_0 - \lambda_1\tilde U_1 \sum_{j=0}^5 e^{i\ell^2\GG_j\times\kk}}
	{\Phi_0 - \lambda_1\Phi_1 \sum_{j=0}^5 e^{i\ell^2\GG_j\times\kk}}.
\end{equation}
According to the expression of $U$ in Eq. (\ref{eq_ideal_pseudopotential}), the flat band condition $\tilde U_0/\Phi_0 = \tilde U_1/\Phi_1$ implies
that the bandwidth vanishes for 
\begin{equation}
	\omega_{c0}^{\rm AC1} = \frac{\Phi_0\tilde\Delta_{+,1} - \Phi_1\tilde\Delta_{+,0}}{\Phi_0\tilde\xi_1 - \Phi_1\tilde\xi_0},
    \label{eq_app_omegac0AC1}
\end{equation}
where $\tilde\Delta_{+,n}$ and $\tilde\xi_{+,n}$ are the $n$th-shell Fourier components of $e^{2\chi(\rr)}\Delta_+(\rr)$ and $e^{2\chi(\rr)}\xi(\rr)$, respectively.

\begin{figure}
	\centering
	\includegraphics[width=0.48\textwidth]{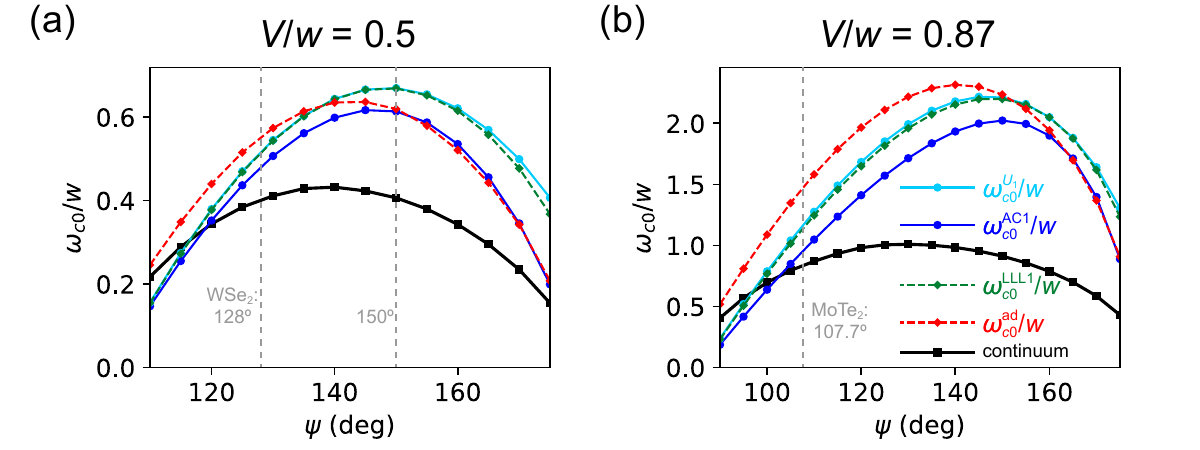}
	\caption{The magic angle, defined as the twist angle with minimum bandwidth, as a function of the model parameter $\psi$, calculated with (a) $V/w = 0.5$ and (b) $V/w = 0.87$ in various approximations including the full continuum model, the full adiabatic approximation ($(N_{\rm LL}, N_{\rm shell}) = (25,15)$), the lowest Landau level projection \cite{morales-duran2024magic}, the AC band projection (Eq. (\ref{eq_app_omegac0AC1})) and $\omega_{c0}^{U_1}$, as indicated by the inset of (b). 
	The $\psi$ values of the MoTe$_2$ and WSe$_2$ models discussed in the main text, as well as the $(V/w, \psi) = (0.5, 150^\circ)$ point, are marked by grey vertical dashed lines.}
	\label{fig_magicangles_app}
\end{figure}

This formula differs quantitatively in two ways 
from the one proposed previously \cite{morales-duran2024magic} based on the $(N_{\rm LL}, N_{\rm shell}) = (1,1)$ limit, 
which gives $\omega_{c0}^{\LLL 1} = \Delta_{+,1}/\xi_1'$ (See Eq. (\ref{eq_LL_U'})). 
First, replacing the AC wave function with the LLL wave function flattens the space-dependent factor $e^{\chi(\rr)}$, leading to $\Phi_1 \rightarrow 0$, $\tilde\xi_1 \rightarrow \xi_1$ and $\tilde\Delta_{+,1} \rightarrow \Delta_{+,1}$, and yields 
$\omega_{c0}^{U_1}$, defined in Sec. \ref{sec_comp_AC_ad}, as the optimal value of $\omega_c$.
Second, the extra term $-\Abs{\mcA'(\rr)}^2/2m$ in Eq. (\ref{eq_LL_U'}),  which takes magnetic field variation into account, 
replaces $\xi_1$ with $\xi_1'$. From Figs. \ref{fig_magicangles_app} (a) and (b), we see that all three ($\omega_{c0}^{\LLL 1}$, $\omega_{c0}^{U_1}$ and $\omega_{c0}^{\rm AC1}$) are generally good approximations to the magic angle of the full adiabatic approximation $\omega_{c0}^{\rm ad}$.
At larger $\psi$'s, $\omega_{c0}^{\rm AC1}$ matches $\omega_{c0}^{\rm ad}$
particularly well, which is another verification of our argument in Appendix \ref{sec_app_res_pot} that the deviation of the full adiabatic approximation from the AC limit is relatively small in this regime due to small higher Fourier components of $U(\rr)$.

\section{Wave Functions of Nearly Ideal Bands in Adiabatic Approximation}
\label{sec_app_ad_id_wf}

\begin{figure}
	\centering
	\includegraphics[width=0.48\textwidth]{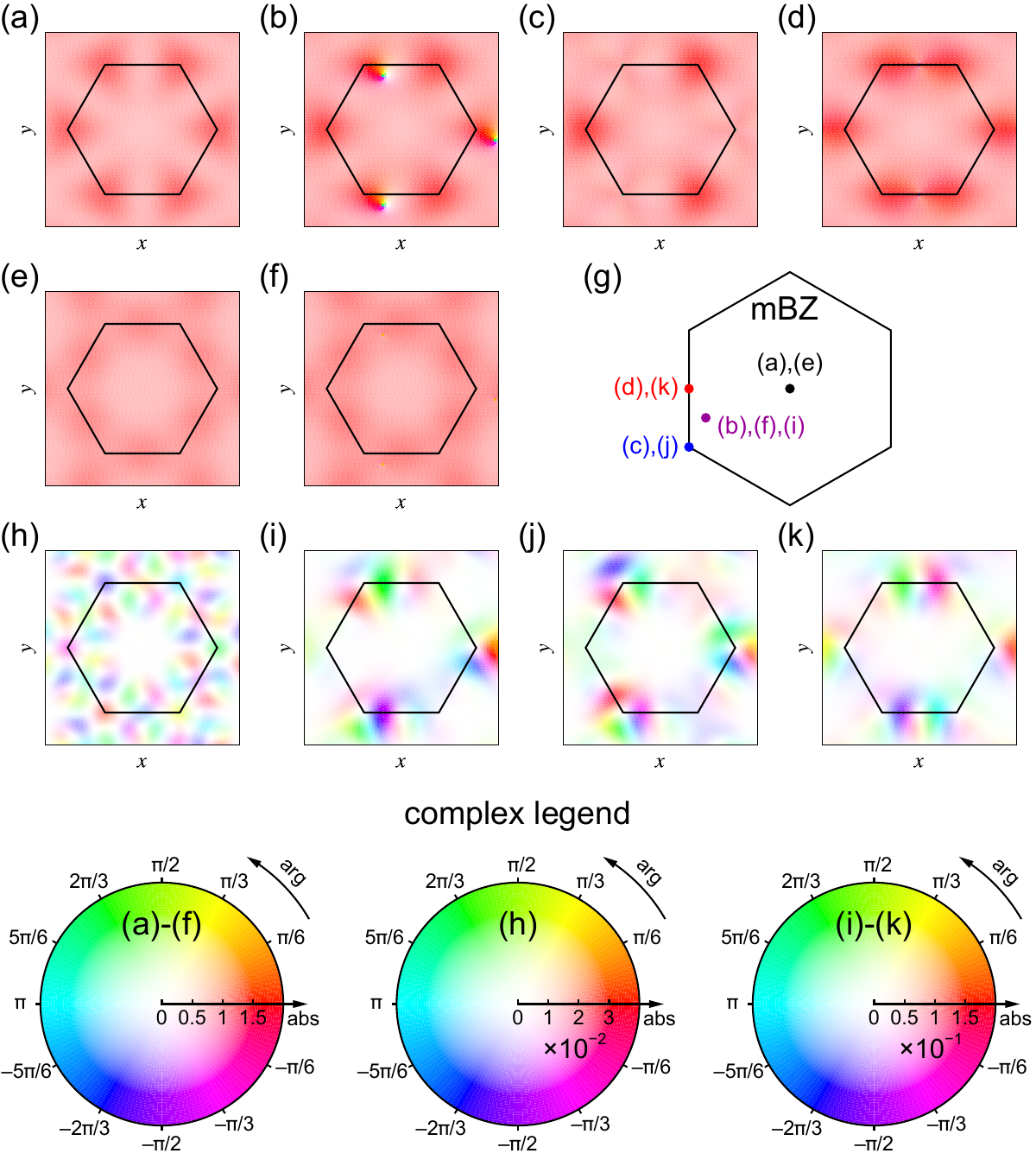}
	\caption{(a)-(d) Plots of the real-space complex functions $\bar\phi_{\kk}(\rr) = \bar\psi_{\kk}^{\rm ad}(\rr)/\psi_{\kk}^{\LLL}(\rr)$ where $\bar\psi_{\kk}^{\rm ad}(\rr)$ is the normalized wave function of the state $\ket{\psi_{\kk}^{\rm ad}}$ calculated under the adiabatic approximation of twisted bilayer \ce{WSe_2} with $\omega_c/w = 0.55$, $(N_{\rm LL}, N_{\rm shell}) = (25, 15)$ and $\kk$ is taken as various points in the mBZ as indicated in (g): (a) $\gamma$ point ($\kk = \0$), (b) some general non-symmetric point $\kk_1$, (c) $k$ point, and (d) $m$ point. (e), (f) The same plots taking $U(\rr) \equiv 0$, \textit{i.e.} the AC limit, but still using the Landau level formulation with $(N_{\rm LL}, N_{\rm shell}) = (25, 15)$, at (e) $\gamma$ point and (f) $\kk_1$. (g) The positions of the $\kk$ points in the mBZ, under which other panels are plotted.
(h)-(k) The plots of $\delta\bar u_{\kk}(\rr)$, which is the component of $\bar u_{\kk}^{\rm ad}(\rr) = e^{-i\kk\cdot\rr}\bar\psi_{\kk}^{\rm ad}(\rr)$ that is orthogonal to $u_{\kk}^{\rm id}(\rr) = e^{-i\kk\cdot\rr} \psi_{\kk}^{\rm id}(\rr)$.
The bottom panels show the color scales used in various complex number maps.
	In (a)-(f) and (h)-(k), the black hexagons are the real-space Wigner-Seitz moir\'e cell.}
	\label{fig_phi}
\end{figure}

We look into the wave functions $\psi_{\kk}^{\rm ad}(\rr)$ of the nearly ideal band identified in the adiabatic approximation of twist bilayer \ce{WSe_2} in Fig. \ref{fig_comparisons} (d) in the main text. Figures \ref{fig_phi} (a)-(d) show that the function $\phi_{\kk}(\rr) = \psi_{\kk}^{\rm ad}(\rr)/\psi_{\kk}^{\LLL}(\rr)$ has about the same shape at all $\kk$ points plotted, which is approximately consistent with the common factor of a 
perfectly ideal $\mcC = 1$ band. Figures \ref{fig_phi} (b)-(d) show that the main difference between $\phi_{\kk}(\rr)$ and $\phi_{\0}(\rr)$ is near the spatial point $-\ell^2\nn_z\times\kk$ (and its translational images), which are the spatial zero points of $\psi_{\kk}^{\LLL}(\rr)$.
For example, the wavefunction ratio $\phi_{\kk_1}(\rr)$ at the low symmetry point $\kk = \kk_1$, illustrated in Fig. \ref{fig_phi} (b),
has a singularity at $-\ell^2\nn_z\times\kk_1$ due to a slight deviation of the 
zero in $\psi_{\kk}^{\rm ad}(\rr)$ from the ideal band zero position.

As a comparison, we reset $U(\rr) = 0$ in the system and recompute this function. Figures \ref{fig_phi} (e)-(f) show perfect
correspondence between $\phi_{\kk}(\rr)$ functions two different $\kk$ points to within numerical accuracy limited by the
truncation of the Hilbert space into 25 Landau levels.
(There is a slight zero shift for $\kk = \kk_1$ case at the zero points of $\psi_{\kk_1}^{\LLL}(\rr)$.) 
The plotted functions accurately approximate the function $e^{\chi(\rr)}$ defined in Sec. \ref{subsec_AC}, which has peaks at the real-space $M$ points due to the potential wells caused by the effective magnetic field pockets in the plasmon analogy. 
The term $U(\rr)$ moves the peaks to the corners of the Wigner-Seitz cell, at which the hole potential wells of $U(\rr)$ sit. These behaviors arise from the same mechanism as the shift in charge density peaks discussed in Sec. \ref{sec_AC_ad_chargedensity}.

To observe the deviation of the wave functions from the ideal limit in the presence of $U(\rr)$,
we define
\begin{equation}
	\Ket{\delta\psi_{\kk}^{\rm ad}} = \Ket{\psi_{\kk}^{\rm ad}} - \frac{\Ket{\psi_{\kk}^{\rm id}} \Braket{\psi_{\kk}^{\rm id}}{\psi_{\kk}^{\rm ad}}}
	{\Braket{\psi_{\kk}^{\rm id}}{\psi_{\kk}^{\rm id}}},
\end{equation}
where $\Ket{\psi_{\kk}^{\rm id}}$ is the
perfect $\mcC = 1$ ideal band constructed from the adiabatic charge density $\rho^{\rm ad}(\rr)$ via
\begin{equation}
    \psi_{\kk}^{\rm id}(\rr) = \phi^{\rm id}(\rr) \psi_{\kk}^{\LLL}(\rr) \propto \sqrt{\rho^{\rm ad}(\rr)} \psi_{\kk}^{\LLL}(\rr)
    \label{eq_app_psi_k^id}
\end{equation}
(We have redefined the notation $\phi^{\rm id}(\rr) = e^{\chi^{\rm id}(\rr)}$. See Eq. (\ref{eq_psi_id}) and around). We plot $\delta u_{\kk}^{\rm ad}(\rr) = e^{-i\kk\cdot\rr} \delta\psi_{\kk}^{\rm ad}(\rr)$ in Figs. \ref{fig_phi} (h)-(k) under the normalization choice that $\Abs{\psi_{\kk}^{\rm ad}(\rr)}^2$ averages to 1. We see that not only are the amplitudes of $\delta u_{\kk}^{\rm ad}(\rr)$ at least one order of magnitude smaller than $\psi_{\kk}^{\rm id}(\rr)$, but the ranges of $\delta u_{\kk}^{\rm ad}(\rr)$ are also in general somewhat small compared to the mBZ and mostly concentrated near the zeros of $\psi_{\kk}^{\LLL}(\rr)$ (which are also zeros of $\psi_{\kk}^{\rm id}(\rr)$).

To roughly understand the behavior of quantum geometry, we start by presenting the following general formulas:
\begin{subequations} \label{eq_app_qgformula} \begin{gather}
	\frac{\tr g_{\kk} + \Omega_{\kk}}{4} = \frac{\Braket{\partial_k u_{\kk}}{\partial_k u_{\kk}}}{\Braket{u_{\kk}}{u_{\kk}}} -
	\frac{\Abs{\Braket{u_{\kk}}{\partial_k u_{\kk}}}^2}{\Braket{u_{\kk}}{u_{\kk}}^2}, \label{eq_app_qgformula+} \\\addlinespace
	\frac{\tr g_{\kk} - \Omega_{\kk}}{4} = \frac{\Braket{\partial_{k^*} u_{\kk}}{\partial_{k^*} u_{\kk}}}{\Braket{u_{\kk}}{u_{\kk}}} -
	\frac{\Abs{\Braket{u_{\kk}}{\partial_{k^*} u_{\kk}}}^2}{\Braket{u_{\kk}}{u_{\kk}}^2}, \label{eq_app_qgformula-}
\end{gather} \end{subequations}
where $\Omega_{\kk}$ is the Berry curvature, $\tr g_{\kk}$ is the trace of the Fubini-Study metric, $\ket{u_{\kk}} = e^{-i\kk\cdot\rr} \ket{\psi_{\kk}}$ is the cell (quasi-) periodic part of the band wave function of a general band, and $k = k_x + ik_y$. In our notation, $\bra{\partial_k u_{\kk}}$ is the adjoint of $\ket{\partial_k u_{\kk}}$, which means that $\bra{\partial_k u_{\kk}} = \partial_{k^*} \bra{u_{\kk}}$. Eqs. (\ref{eq_app_qgformula}) hold for arbitrary $\kk$-space gauge, including the unnormalized ones.

We next apply these formulas to our adiabatic approximation band where
\begin{equation}
	\Ket{u_{\kk}^{\rm ad}} = e^{-i\kk\cdot\rr} \Ket{\psi_{\kk}^{\rm ad}} = \Ket{u_{\kk}^{\rm id}} + \Ket{\delta u_{\kk}^{\rm ad}}, \quad
	\Braket{u_{\kk}^{\rm id}}{\delta u_{\kk}^{\rm ad}} = 0.
\end{equation}
We first choose the gauge so that the ideal band $\ket{u_{\kk}^{\rm id}}$ is holomorphic in $k$ (\textit{i.e.} has no $k^*$ derivative) and apply Eq. (\ref{eq_app_qgformula-}). From $\ket{\partial_{k^*} u_{\kk}^{\rm ad}} = \ket{\partial_{k^*} \delta u_{\kk}^{\rm ad}}$ it is clear that $\tr g_{\kk} - \Omega_{\kk}$ is second order in $\ket{\partial_{k^*} \delta u_{\kk}^{\rm ad}}$, giving the extremely ideal quantum geometry. Then we select another gauge where $u_{\kk}^{\rm id}(\rr) = \phi^{\rm id}(\rr) u_{\kk}^{\LLL}(\rr)$ (from Eq. (\ref{eq_app_psi_k^id})) with no $\kk$ dependence further than the normalized cell-quasiperiodic LLL function $u_{\kk}^{\LLL}(\rr) = e^{-i\kk\cdot\rr} \psi_{\kk}^{\LLL}(\rr)$. Because of orthogonality, the contribution of $\ket{\delta u_{\kk}^{\rm ad}}$ to the denominators of all terms of Eq. (\ref{eq_app_qgformula+}) is second order, thus negligible. We also assume that $\tr g_{\kk} \approx \Omega_{\kk}$, hence we have
\begin{widetext} \begin{equation}
	\frac{\Omega_{\kk}^{\rm ad} - \Omega_{\kk}^{\rm id}}{4} \approx
	\frac{\Braket{\partial_k u_{\kk}^{\rm id}}{\partial_k \delta u_{\kk}^{\rm ad}}}{\Braket{u_{\kk}^{\rm id}}{u_{\kk}^{\rm id}}} -
	\frac{\Braket{\partial_k u_{\kk}^{\rm id}}{u_{\kk}^{\rm id}} \Braket{u_{\kk}^{\rm id}}{\partial_k \delta u_{\kk}^{\rm ad}}}
	{\Braket{u_{\kk}^{\rm id}}{u_{\kk}^{\rm id}}^2} + c.c..  \label{eq_app_Omegadif_adid}
\end{equation} \end{widetext}
Though both terms in Eq. (\ref{eq_app_Omegadif_adid}) are first order in $\ket{\partial_k \delta u_{\kk}^{\rm ad}}$, the function $\partial_k \delta u_{\kk}^{\rm ad}(\rr)$ tends to localize near the zero points of $u_{\kk}^{\rm id}(\rr)$ as we have observed in Fig. \ref{fig_phi} (i)-(k), further suppressing the second term. On the other hand, $\partial_k u_{\kk}^{\rm id}(\rr)$ tends to have large absolute values near the zero points of $u_{\kk}^{\rm id}(\rr)$, making the first term relatively significant. Furthermore, since $\partial_k u_{\kk}^{\rm id}(\rr)$ contains a $\kk$-independent factor $\phi^{\rm id}(\rr)$, which has peaks at the corners of the Wigner-Seitz cell, the inner product on the numerator of the first term tends to be larger at those $\kk$ points where $\partial_k \delta u_{\kk}^{\rm ad}(\rr)$ are localized near the Wigner-Seitz cell corners, which are the mBZ corners. In the meantime, the denominator of the first term cannot have significant peaks at the mBZ corners due to suppression from the magnetic form factor $\lambda_{\GG}$ when applying the formula Eq. (\ref{eq_Berrycurv_ACb}) to the ideal band. These observations explain the occurrence of the Berry curvature peaks at the mBZ corner upon deviation from the ideal band limit.

\section{Relation Between Charge Density and Common Factor of Ideal Bands}
\label{sec_app_rho_Phi}

Under the normalization convention we define in Sec. \ref{subsec_AC}, the charge density of a full ideal band with wave function $\psi_{\kk}^{\rm id}(\rr)$ specified in Eq. (\ref{eq_psi_id}) is
\begin{equation} \begin{array}{c}
    \rho^{\rm id}(\rr) = \frac{1}{A} \sum_{\kk\in\rm mBZ} \frac{\Abs{\psi_{\kk}^{\rm id}(\rr)}^2}{\Braket{\psi_{\kk}^{\rm id}}{\psi_{\kk}^{\rm id}}} \\\addlinespace
    = \frac{e^{2\chi^{\rm id}(\rr)}}{A} \sum_{\kk\in\rm mBZ} \frac{\Abs{\psi_{\kk}^{\LLL}(\rr)}^2}{\Braoperket{\psi_{\kk}^{\LLL}}{e^{2\chi^{\rm id}(\rr)}}{\psi_{\kk}^{\LLL}}},
\end{array} \label{eq_app_rho_id} \end{equation}
where $A$ is the total system area. The Fourier components of the spatial function $|\psi_{\kk}^{\LLL}(\rr)|^2$ can be obtained from Eqs. (\ref{eq_app_mcI}) and (\ref{eq_app_mcIresult}):
\begin{equation}
    \frac{1}{A} \int d^2\rr e^{-i\GG\cdot\rr} \Abs{\psi_{\kk}^{\LLL}(\rr)}^2 = \mcI_{\kk\kk}^{-\GG} = \overline\eta_{\GG} \lambda_{\GG} e^{i\ell^2\kk\times\GG},
\end{equation}
and the denominator in the summation in Eq. (\ref{eq_app_rho_id}) can be directly copied from the AC case presented in Eq. (\ref{eq_Berrycurv_ACb}), which then gives
\begin{equation}
    \rho^{\rm id}(\rr) = \frac{e^{2\chi^{\rm id}(\rr)}}{A} \sum_{\kk\in\rm mBZ}
    \frac{\sum_{\GG} \overline\eta_{\GG} \lambda_{\GG} e^{i\left( \GG\cdot\rr + \ell^2\kk\times\GG \right)}}{\sum_{\GG} \overline\eta_{\GG} \lambda_{\GG} \Phi_{\GG}^{\rm id} e^{i\ell^2\GG\times\kk}},
    \label{eq_app_rho_id_1}
\end{equation}
where $\Phi_{\GG}^{\rm id}$ is the Fourier component of $e^{2\chi^{\rm id}(\rr)}$.

Now we apply the first-shell approximation to both the numerator and the denominator in the summation of Eq. (\ref{eq_app_rho_id_1}) as we did in Appendix \ref{sec_app_AClimit}, and then do a Taylor expansion over $\lambda_1$:
\begin{equation} \begin{array}{c}
    \rho^{\rm id}(\rr) \approx \frac{e^{2\chi^{\rm id}(\rr)}}{A\Phi_0^{\rm id}} \sum_{\kk\in\rm mBZ}
    \left( 1 - \lambda_1 \sum_{j=0}^5 e^{i \left( \GG_j\cdot\rr + \ell^2\kk\times\GG_j \right)} \right)
    \\\addlinespace
    \times \left( 1 + \frac{\lambda_1 \Phi_1^{\rm id}}{\Phi_0^{\rm id}} \sum_{j'=0}^5 e^{i\ell^2\GG_{j'}\times\kk} + \mathscr{O}\left(\lambda_1^2\right) \right).
\end{array} \end{equation}
Since all nontrivial harmonic terms in $\kk$ (\textit{i.e.} that depend on $\kk$ as $\sim e^{i\ell^2\kk\times\GG}$ with nonzero reciprocal lattice vector $\GG$) are integrated out by summation over the mBZ, up to the leading term of $\rr$ dependence we get
\begin{equation}
    \rho^{\rm id}(\rr) \approx \frac{e^{2\chi^{\rm id}(\rr)}}{A_M \Phi_0^{\rm id}} \left( 1 - \frac{\lambda_1^2 \Phi_1^{\rm id}}{\Phi_0^{\rm id}} \sum_{j=0}^5 e^{i\GG_j\cdot\rr} \right).
    \label{eq_app_rho_id_2}
\end{equation}
We see that compared to $e^{2\chi^{\rm id}(\rr)}$, the spatially varying part of the bracket in Eq. (\ref{eq_app_rho_id_2}) is suppressed by a factor of $\lambda_1^2 \approx 0.027$, hence conclude that $\rho^{\rm id}(\rr)$ has almost identical shape as $e^{2\chi^{\rm id}(\rr)}$.

\bibliography{bibliography}

\end{document}